\documentclass[preprint,pteplogo]{ptephy_v2}
\preprintnumber{XXXX-XXXX} 
\usepackage{hyperref}

\begin{document}

\title{An exact Markovian decoherence dynamics  of two interacting harmonic oscillators coupled to a bosonic heat bath}

\author{G\'abor Homa}
\affil{ HUN-REN Wigner Research Centre for Physics, Konkoly-Thege M. \'ut 29-33, H-1121 Budapest, Hungary \email{homa.gabor@wigner.hun-ren.hu}}

\author{D\'avid Hamar}
\affil{Department of Pharmaceutics, Semmelweis University, 1092 Budapest, Hungary \email{hamar.david@semmelweis.hu}}

\author{J\'ozsef Zsolt Bern\'ad}
\affil{ HUN-REN Wigner Research Centre for Physics, Konkoly-Thege M. \'ut 29-33, H-1121 Budapest, Hungary}
\affil{Forschungszentrum J\"ulich, Institute of Quantum Control (PGI-8), 52425 J\"ulich,
Germany \email{j.bernad@fz-juelich.de} }

\author{Peter Adam}
\affil{ HUN-REN Wigner Research Centre for Physics, Konkoly-Thege M. \'ut 29-33, H-1121 Budapest, Hungary \email{adam.peter@wigner.hun-ren.hu}}
\affil{Institute of Physics, University of P\'ecs, Ifj\'us\'ag \'utja 6, H-7624 P\'ecs, Hungary}

\author{Andr\'as Csord\'as}
\affil{Department of Physics of Complex Systems, E\"{o}tv\"{o}s Lor\'and University, ELTE, P\'azm\'any P\'eter s\'et\'any 1/A, H-1117 Budapest, Hungary \email{csordas@tristan.elte.hu}}

\begin{abstract}
We present an exact analytical solution of the Hu-Paz-Zhang master equation in a precise Markovian limit for a system of two harmonically coupled harmonic oscillators interacting with a common thermal bath of harmonic oscillators. The thermal bath is initially considered to be at arbitrary temperatures and characterized by an Ohmic Lorentz-Drude spectral density. In the examined system, couplings between the two harmonic oscillators and the environment ensure a complete decoupling of the center-of-mass and relative degrees of freedom, resulting in undamped dynamics in the relative coordinate. The exact time evolution is used to analyze the system's entanglement dynamics, quantified through logarithmic negativity and quantum mutual information, while ensuring the positivity of the density operator to confirm the physical validity of the results. We demonstrate that, under certain parameter regimes and initial conditions, the asymptotic dynamics can give rise to periodic entanglement-disentanglement behavior. Furthermore, numerical simulations reveal that for negative values of the direct coupling between the oscillators, which are sufficiently close to a critical lower bound beyond which the system becomes unstable, the system can maintain entanglement across a broad temperature range and for arbitrarily long durations.
\end{abstract}

\subjectindex{A60, A61, A64}

\maketitle

\section{Introduction}

The investigation of coupled harmonic oscillators constitutes a fundamental aspect of contemporary physics, owing to their wide applicability in diverse fields such as quantum optomechanics \cite{Groblacher2009}, biophysics \cite{Romero2014, Halpin2014}, quantum chemistry \cite{Thomas2018}, cavity QED \cite{Brown2011,Ding2017}, superconducting circuits \cite{Bothner2021}, and quantum technology \cite{Rigatos2009}. These experimental efforts have been supported by a substantial body of theoretical research. The prototype model has proven versatile in addressing a wide array of phenomena, including deep inelastic collisions in nuclear physics \cite{Sandulescu_1987}, entanglement between two harmonic oscillators \cite{RajagopalPRA2001, Dodonov_2005, Benatti_2006, Paz2008, Galve, Isar_2013, AbbasnezhadQIP2017, MakarovPRE2018, Volkoff, Chapman, MakarovMath2025}, coherence dynamics \cite{dePonteAoP2005, Chou}, synchronization processes \cite{HenrietPRA2019, GargPLA2023, BittnerJCP2025}, and quantum metrology \cite{Huang}. The literature on these topics is extensive, even without considering foundational works such as the study of parametric amplification in quantum optics \cite{Mollow}. 

In addition to direct interactions, quantum mechanical systems can also influence each other indirectly through a shared environment that mediates their mutual interaction.
Some of the aforementioned studies have already begun exploring this direction \cite{Benatti_2006, Chou, Paz2008, Isar_2013, HenrietPRA2019, BittnerJCP2025}. Mediated interactions also play a significant role in other areas of physics, such as cooperative spontaneous emission \cite{Dicke, Stephen, Milonni}. In a broader sense, models in which quantum systems are coupled to a common environment serve as paradigmatic examples of open quantum systems. Their investigation is crucial not only for harnessing quantum effects in emerging technologies such as lasers and masers \cite{Scully}, inhomogeneous spin ensembles \cite{Staudt}, and quantum computing architectures \cite{Majidy}, but also for deepening our understanding of how classical behavior emerges through decoherence \cite{Joos}. A general challenge is that the master equations governing these open quantum systems cannot be solved exactly. Consequently, approximation-induced nonphysical behavior, including the loss of positivity of the density operator during time evolution, cannot be properly addressed. Therefore, in this work, we study a system of two harmonically coupled oscillators interacting with a common thermal bath composed of harmonic oscillators. The single harmonic oscillator case has been extensively studied, yielding the exact Hu-Paz-Zhang (HPZ) master equation \cite{HPZ, H-Y, book1, Fleming}. This belongs to the class of time-convolutionless master equations and can be viewed as a particular case of the Nakajima-Zwanzig equation \cite{Nakajima, Zwanzig}. For various applications, physicists often approximate the non-Markovian HPZ master equation by a Markovian counterpart. However, such simplifications can introduce inconsistencies, e.g. positivity violations of the density operator during the time evolution, which we have already pointed out some years ago \cite{BLH, HBCSCS}. Accordingly, our goal is to obtain an exact solution of the master equation describing a system of two coupled harmonic oscillators in the Markovian limit and to analyze its dynamics.

Focusing on the Markovian limit, we begin by deriving the general solution using the method of characteristics \cite{Courant}. This approach exploits the fact that the system can be mapped onto an equivalent model comprising a single harmonic oscillator coupled to a thermal bath, along with an uncoupled free harmonic oscillator \cite{Chou}. First, we demonstrate that in the asymptotic limit, the system does not approach a stationary state; specifically, certain oscillatory components of the density operator persist undamped as $t \to \infty$. Second, we establish additional criteria to ensure the positivity of the time-evolved quantum state. We then consider a thermal bath characterized by a Lorentz-Drude Ohmic spectral density. Building on our previous work \cite{HBCS}, this choice allows for the exact analytic derivation of the Markovian coefficients in the HPZ master equation. While formal expressions for these coefficients are known (see, e.g., the work of Halliwell and Yu \cite{H-Y}), their evaluation is typically performed either perturbatively \cite{Fleming, Breuer} or using less realistic spectral densities \cite{Paz2008} and inconsistent formulas \cite{Ford}. Finally, we analyze the system within the framework of continuous-variable quantum information theory \cite{Lami,japcsi}, complemented by a consistency check of the underlying master equations.

 The paper is organized as follows. In Sec.~\ref{sec:II}, we analytically solve the general HPZ master equation in the Markovian limit using the method of characteristics. In this section, we also define and compute the mathematical stability criteria and the positivity criteria that govern the asymptotic physical behavior of the solution to the Markovian HPZ master equation. Subsequently, we derive the exact Markovian coefficients of the master equation with a Lorentz-Drude type Ohmic spectral density at arbitrary temperatures and classify these coefficients according to all relevant physical cases. In Sec.~\ref{sec:III}, we present a brief theoretical summary of quantum mutual information and logarithmic negativity and apply these measures to Gaussian density operators obtained during the evolution of the Markovian HPZ master equation. In Sec.~\ref{sec:IV}, we examine several scenarios that may be of interest with respect to the Markovian decoherence dynamics of entanglement between two oscillators.
Finally, Sec.~\ref{sec:V} summarizes the results and provides concluding remarks along with a brief outlook.

\section{Two coupled oscillators interacting with an oscillator bath} \label{sec:II}
\subsection{The master equation and its general Markovian solution } \label{sec:master_eq}

Let us consider a quantum mechanical system of two oscillators with equal masses $m$ and oscillator frequencies $\Omega$ coupled to each other harmonically with coupling constant $\kappa$:
\begin{equation}
		\hat{H}_\text{sys}=\frac{\hat{P}_1^2}{2m}+\frac{1}{2}m\Omega^2\hat{x}_1^2+\frac{\hat{P}_2^2}{2m}+\frac{1}{2}m\Omega^2\hat{x}_2^2+\kappa(\hat{x}_1-\hat{x}_2)^2. \label{Hamiltonian_sys}
\end{equation}
Initially, the thermal bath at temperature $T$ consists of a set of non-interacting oscillators with masses $m_n$ and bath oscillator frequencies $\omega_n$:
\begin{equation}
\hat{H}_\text{bath}=\sum_n \left(\frac{\hat{p}_n^2}{2m_n}+\frac12 m_n\omega_n^2 \hat{q}_n^2 \right).
\end{equation}
Let us suppose that the two central oscillators are coupled to the $n$-th bath oscillators also harmonically with coupling constants $C_n$ through the center-of-mass (CM) coordinate of the two central oscillator as follows
\begin{equation}
\hat{H}_\text{int}=\frac{(\hat{x}_1+\hat{x}_2)}{2}\sum_n C_n \hat{q}_n.
\label{eq:interaction_term_in_H}
\end{equation}
In the literature, sometimes the convention $C_n \to 2C_n$ is used \cite{Chou}.
The Hamiltonian for the total system is:
\begin{equation} \hat{H}=\hat{H}_\text{sys}+\hat{H}_\text{bath}+\hat{H}_\text{int}.
\label{Hamiltonian_full}
\end{equation}
$\hat{x}_i,\hat{q}_n$ and $\hat{P}_i,\hat{p}_n$ are position and momentum operators with the usual commutators. The same model has been considered in Ref.~\cite{Chou} (See this paper and references therein). 

The form of the coupling suggests to introduce position and momentum operators belonging to the center-of-mass and relative (CMR) degrees of freedom:
\begin{eqnarray}
		\hat{X}&:=&\frac{1}{2}(\hat{x}_1+\hat{x}_2), \quad \hat{P}:=\hat{P}_1+\hat{P}_2, \quad M:=2m \nonumber \\
		\hat{x}&:=&\hat{x}_1-\hat{x}_2, \quad \hat{p}:=\frac{1}{2}(\hat{P}_1-\hat{P}_2), \quad \nonumber \mu:=\frac{m}{2}, \nonumber \\
        \hat{\mathbf{w}}&:=&(\hat{X},\hat{P},\hat{x},\hat{p})^\top.
\end{eqnarray} 
We will refer to this set of new operators as the CMR operators. The Hamiltonian expressed in terms of the CMR operators takes the form:
\begin{eqnarray}
		\hat{H}=\underbrace{\frac{\hat{P}^2}{2M}+\frac{1}{2}M\Omega^2\hat{X}^2}_{=\hat{H}_\text{cm}}+\underbrace{\frac{\hat{p}^2}{2\mu}+\frac{1}{2}\mu\Omega^2\hat{x}^2+\kappa\hat{x}^2}_{=\hat{H}_\text{rel}} 
         +  \hat{X}\sum_n C_n \hat{q}_n +\sum_n \left(\frac{\hat{p}_n^2}{2m_n}+\frac12 m_n\omega_n^2 \hat{q}_n^2 \right). \label{Hamiltonian_CM}
\end{eqnarray}
The evolution of the total system’s state, represented by the density operator $\hat{\rho}_\text{total}(t)$, is governed by the von Neumann equation
\begin{equation}
    \frac{d\hat{\rho}_{\text{total}}(t)}{dt}=-\frac{i}{\hbar}\left [\hat{H},\hat{\rho}_{\text{total}}(t)\right].
\end{equation}
By eliminating the bath degrees of freedom, the state of the two harmonic oscillators is described by $\hat{\rho}(t)$, which is obtained by tracing over the bath, i.e. $\hat{\rho}(t)=\mathrm{Tr}_\text{bath}\{\hat{\rho}_{\text{total}}(t)\}$. 
We assume that the thermal bath and the two oscillators are initially uncorrelated. This implies that the initial state of the  CM oscillator, described by the Hamiltonian $\hat{P}^2/(2M) + M\Omega^2 \hat{X}^2/2$, is separable from the thermal bath. As a result, the CM harmonic oscillator follows the HPZ master equation. The Hamiltonian $\hat{H}_\text{sys}$ includes an additional term that depends only on the relative operators $\hat{p}$ and $\hat{x}$. The full dynamics of the two harmonic oscillators is described by
\begin{eqnarray} 
			\frac{\partial \hat{\rho}(t)}{ \partial t}&=&-\frac{i}{\hbar}\left [\hat{H}_\text{cm}+\hat{H}_\text{rel},\hat{\rho}(t)\right]+ \frac{A(t)}{2 i\hbar}[\hat{X}^2,\hat{\rho}(t)] 
			+\frac{B(t)}{2 i \hbar} [\hat{X},\{ \hat{P},\hat{\rho}(t)\}]\nonumber \\
			&& +\frac{C(t)}{ \hbar^2}[\hat{X},[\hat{P},\hat{\rho}(t)]] - \frac{D(t)}{ \hbar^2}[\hat{X},[\hat{X},\hat{\rho}(t)]] , \label{eq:HPZ_comm}
		\end{eqnarray}
where $[,]$ stands for commutators, while $\{,\}$ for anti-commutators and  $A(t), B(t), C(t)$ and $D(t)$ are the non-Markovian time-dependent coefficients describing the influence of  environment of the bath oscillators.

In the next step, we consider two representations. The one in the original phase space coordinates is defined as
\begin{eqnarray}
		\rho(\Delta_1,k_1,\Delta_2,k_2, t)=
        \mathrm{Tr} \big\{\hat{\rho}(t) \exp\{ik_1 \hat{x}_1- i \Delta_1 \hat{P}_1+ik_2 \hat{x}_2- i \Delta_2 \hat{P}_2\} \big\} \label{eq:rho_orig}, 
\end{eqnarray}
and	the second one defined in terms of the CMR operators
\begin{equation}
		\rho_c(\Delta,K,\delta,k, t)= \mathrm{Tr} \big\{\hat{\rho}(t) \exp\{ik \hat{x}- i \delta \hat{p}+iK \hat{X}- i \Delta \hat{P}\} \big\},\label{rho_in_cm}
\end{equation}
where
\begin{eqnarray}
		K&:=&k_1+k_2, \quad k:=\frac{1}{2}(k_1-k_2), \nonumber \\
        \Delta&:=&\frac{1}{2}(\Delta_1+\Delta_2), \quad \delta:=\Delta_1-\Delta_2. 
        \end{eqnarray}
Finally, after performing the necessary transformations and calculations, we arrive at the following equation:
\begin{eqnarray}
 \displaystyle\frac{\partial}{\partial t} \rho_c(\mathbf{w},t) &=&  \biggl[-\frac{\hbar K}{M} \frac{\partial}{\partial \Delta}-\frac{\hbar k}{\mu} \frac{\partial}{\partial \delta}+ 
\frac{M \Omega^2}{\hbar}  \Delta\frac{\partial}{\partial K}       
+ \frac{\mu\Omega^2+2\kappa}{\hbar} \delta \frac{\partial}{\partial k}   
    \nonumber \\
&&\quad +  \frac{A(t)}{\hbar} \Delta \frac{\partial}{\partial K} -  B(t)  \Delta \frac{\partial}{\partial \Delta} \quad +\frac{C(t)}{\hbar} \Delta K - \frac{D(t)}{\hbar^2} \Delta^2 
			\biggr]\rho_c(\mathbf{w},t),  \label{eq:master_equation}
\end{eqnarray}
where 
\begin{equation} \label{eq:CM_vect}
    \mathbf{w}^\top=(\Delta, K, \delta, k).
\end{equation}
 The initial condition to Eq. \eqref{eq:master_equation} is given by  
 \begin{equation} \label{eq:initial_value}
\rho_c(\mathbf{w},0)= \rho_{c_0} (\mathbf{w})  .
 \end{equation}
We note that $\mathrm{Tr}\,\hat{\rho}=1$ requires
\[
 \rho_{c_0} (\mathbf{w}=\mathbf{0})=1.
\]
As the notation suggests, it is advantageous to consider $\rho_{c_0}$ as a function with vector-dependence $\mathbf{w}$.
Let us introduce the following quantities:
 \begin{eqnarray}
 \omega^2_c&:=&\Omega^2+\frac{A}{2m}, \quad \omega^2_d:=\Omega^2+\frac{4 \kappa}{m}, \quad B:=2\lambda, \quad
 C:=2\hbar D_{px}, \quad \text{and}\quad D:=\hbar D_{pp}.\quad
 \label{eq: new_frequencies}
 \end{eqnarray}
Then, in the Markovian limit $A(t)$, $B(t)$, $C(t)$ and $D(t)$, and correspondingly $\omega_c$, $\omega_d$, $\lambda$, $D_{px}$  and $D_{pp}$ tend to constant, and   (\ref{eq:master_equation}) can be written as
\begin{eqnarray}
  && \displaystyle\frac{\partial \rho_c}{\partial t}  +\left( \frac{\hbar K}{2m}+2\lambda \Delta\right)\frac{\partial \rho_c}{\partial \Delta}-\frac{2m \omega_c^2 \Delta}{\hbar} \frac{\partial \rho_c}{\partial K}+\frac{2\hbar k}{m}\frac{\partial \rho_c}{\partial \delta}
 -\frac{m\omega_d^2 \delta}{2\hbar} \frac{\partial \rho_c}{\partial k} 
 \nonumber \\
  && \quad \quad =\left(2D_{px} \Delta K -\frac{D_{pp}}{\hbar} \Delta^2 \right)\rho_c.
 \label{eq:master_eq_rewritten}
\end{eqnarray}
The left-hand-side is a total derivative, if one identifies the following equations for the characteristics:
\begin{eqnarray}
\frac{ d\Delta}{dt}&=&\frac{\hbar K}{2m}+2\lambda \Delta, \quad \frac{ d K}{dt}=-\frac{2m \omega_c^2 \Delta}{\hbar},  \frac{ d\delta}{dt}= \frac{2\hbar k}{m}, \quad \frac{ d k}{dt}=-\frac{m\omega_d^2 \delta}{2\hbar}.
\label{eq:ODE_for_characterteristics}
\end{eqnarray}
After those identifications, one has to solve from Eq. (\ref{eq:master_eq_rewritten})
\begin{eqnarray}
\frac{d \rho_c}{d t}= -\left(\frac{D_{pp}}{\hbar} \Delta^2 (t)-2D_{px} \Delta(t) K(t)\right) \rho_c(t) \equiv -E(t)\rho_c(t),
\label{eq:ODE_for_rho}
\end{eqnarray}
where the time dependence of $\Delta(t)$ and $K(t)$ is taken from the solutions of 
(\ref{eq:ODE_for_characterteristics}).
The equations (\ref{eq:ODE_for_characterteristics}) are linear in the phase-space variables with a constant coefficient matrix 
 \begin{eqnarray}
\frac{d}{dt}  \begin {pmatrix}\Delta\\ \noalign{\medskip}K
\\ \noalign{\medskip}\delta\\ \noalign{\medskip}k\end {pmatrix}=     \begin {pmatrix} 2 \lambda &\frac{\hbar}{2m}&0&0\\ \noalign{\medskip}
-\frac{2 m{\omega_c}^{2}}{\hbar}&0&0&0\\ \noalign{\medskip}0&0&0&\frac{2 \hbar}{m}
\\ \noalign{\medskip}0&0&-\frac{m{\omega_d}^{2}}{2\hbar}&0
\end {pmatrix} 
          \begin {pmatrix}\Delta\\ \noalign{\medskip}K
\\ \noalign{\medskip}\delta\\ \noalign{\medskip}k\end {pmatrix} \equiv \mathbf{M} \cdot\begin{pmatrix} \Delta \\K \\ \delta \\k \end{pmatrix}.\label{eq:characteristic_equation}
 \end{eqnarray}
The solution of (\ref{eq:characteristic_equation}) is
\begin{equation}
\mathbf{w}(t)=\exp{(\mathbf{M} t)}\cdot \mathbf{w}(0).
\label{eq:evolution_of characteristics}
\end{equation}
The non-symmetric matrix $\mathbf{M}$ has constant elements and can be decomposed as
\[
\mathbf{M}=\sum_{j=1}^4 \Lambda_j \mathbf{P}_j,
\]
where the $\Lambda_i$'s are the eigenvalues of  $\mathbf{M}$ and the matrices $\mathbf{P}_i$ are mutually orthogonal projectors  that sum to the $4 \times 4$ identity matrix, i.e., 
\begin{equation} \label{eq:proj_identities}
\mathbf{P}_j\mathbf{P}_k=\delta_{j,k}\mathbf{P}_j, \quad \sum_{j=1}^4 \mathbf{P}_j=I_4.
\end{equation}
The projectors $\mathbf{P}_i$ can be calculated from the left and right eigenvectors. In our case, the eigenvalues are
\begin{equation}
\Lambda_{1,2}=\lambda\pm\sqrt{\lambda^2-\omega^2_c}   \quad  \text{and } \Lambda_{3,4}= \pm i \omega_d, 
\end{equation}
and the projectors are
\begin{eqnarray}
 \mathbf{P}_j &=& \frac{1}{1-\frac{\Lambda^2_j}{\omega^2_c}}     \begin {pmatrix} -\frac{\Lambda^2_j}{\omega^2_c} &-\frac{\Lambda_j}{2m  \omega^2_c}&0&0
\\ \noalign{\medskip}2m\Lambda_j&1&0&0
\\ \noalign{\medskip}0&0&0&0\\ \noalign{\medskip}0&0&0
&0\end {pmatrix}, \quad  \text{for} \quad j=1,2
\nonumber \\
\mathbf{P}_j&=&\begin{pmatrix}
0 & 0 & 0 & 0 \\
0 & 0 & 0 & 0 \\
0 & 0 & \frac12 & -\frac{  \hbar }{m \Lambda_j} \\
0 & 0 & -\frac{ m \Lambda_j }{4\hbar} & \frac12
\end{pmatrix},\quad \text{for} \quad j=3,4.
\end{eqnarray}
It will turn out that the stability will require that $\mathrm{Re}\,\Lambda_j \geq 0$ for all $j$, which leads to 
\begin{equation}
\lambda \geq 0, \quad \omega_c \geq 0 , \quad \omega_d^2 \geq 0.
\label{eq:stability_criterions}
\end{equation}
These requirements give bounds for $A$ and for the coupling $\kappa$:
\[
-2m\Omega^2 \leq A, \quad - \frac{m\Omega^2}{4} = \kappa_{\text{crit}} \leq \kappa.
\]
Using projectors $\mathbf{P}_k$, one has a useful form for  evolution (\ref{eq:evolution_of characteristics}) 
\begin{equation}
\mathbf{w}(t)=\exp{(\mathbf{M} t)}\cdot \mathbf{w}(0)=\sum^4_{k=1} \exp\left({\Lambda_k t}\right) \mathbf{P}_k \cdot  \mathbf{w}(0).
\label{eq:characteristic_curves}
\end{equation}

The next step is to solve (\ref{eq:ODE_for_rho}), whose solution is
\begin{equation}
\rho_c(t)=\rho_c(0) \exp\left( -\int_0^t E(s) \, ds\right),
\end{equation}
which is considered as
\begin{equation} \label{eq: rho_c}
\rho_c(\mathbf{w}(t),t)=\rho_c(\mathbf{w}(0),0) \exp(f(\mathbf{w}(0),t)),
\end{equation}
where the exponential factor should be expressed in terms of $\mathbf{w}(0)$ and $t$, and
\[
f(\mathbf{w}(0),t)= -\int_0^t E(s) \, ds.
\]
Trivially $f(\mathbf{w}(0),0)=0$, thus, by the initial condition \eqref{eq:initial_value}
\[
\rho_c(0) \equiv \rho_c(\mathbf{w}(0))=\rho_{c_0}(\mathbf{w}(0)).
\]
$\mathbf{w}(t)$ and $\mathbf{w}(0)$ are connected by the characteristic curves (\ref{eq:characteristic_curves}).
Let us introduce the matrix $\mathbf{R}$ by
\begin{equation}
\mathbf{R}=\begin{pmatrix}
\frac{D_{pp}}{\hbar} & -D_{px} & 0 & 0 \\
-D_{px} & 0 & 0 & 0 \\
0 & 0 & 0 & 0 \\
0 & 0 & 0 & 0 
\end{pmatrix},
\end{equation}
and let us write $E(s)$ as
\begin{eqnarray}
E(s)&=&\left(\frac{D_{pp}}{\hbar} \Delta^2 (s)-2D_{px} \Delta(s) K(s)\right)
=\mathbf{w}(s)^\top \cdot \mathbf{R} \cdot \mathbf{w}(s) \nonumber\\ 
&=& \mathbf{w}(0)^\top \cdot  \!\!
\sum_{k,l=1}^{4} \!\! \exp(\Lambda_k s) \mathbf{P}_k^\top \cdot \mathbf{R} \cdot \exp(\Lambda_\ell s)\mathbf{P}_\ell \cdot \mathbf{w}(0). \nonumber
\end{eqnarray}
Here we used (\ref{eq:characteristic_curves}) to express $\mathbf{w}(s)$ with $\mathbf{w}(0)$. Due to the block in the upper left corner of $\mathbf{R}$ only the $k,\ell=1,2$ terms give a contribution to $E(s)$.  Integration is trivial, and we get
for $f(\mathbf{w}(0),t)$
\begin{eqnarray}
 f(\mathbf{w}(0),t)&=&\mathbf{w}(0)^\top \cdot \mathbf{R}_1(t) \cdot \mathbf{w}(0), \nonumber \\ 
 \mathbf{R}_1(t)&=& \sum^2_{k,\ell=1}\mathbf{P}^\top_k \cdot \mathbf{R}\cdot \mathbf{P}_\ell 
 \frac{\exp\left((\Lambda_k+\Lambda_\ell)t\right)-1}{\Lambda_k+\Lambda_\ell} .
\end{eqnarray}
Our main goal is to express $\rho(t)$ as a function of $t$ and the phase space coordinate $\mathbf{w}$ (independent of $t$), which requires special choices of initial conditions for which the trajectory $\mathbf{w}(s)$ ends here at time $t$:
\[
\mathbf{w}=\mathbf{w}(t)=\exp(\mathbf{M} t) \cdot \mathbf{w}(0).
\]
It follows that $\mathbf{w}(0)$ should be chosen as 
\begin{equation}
\mathbf{w}(0)=\exp(-\mathbf{M} t) \cdot \mathbf{w} =\sum^4_{k=1} \exp\left({-\Lambda_k t}\right) \mathbf{P}_k \cdot  \mathbf{w}.
\end{equation}
After inserting this $\mathbf{w}(0)$ back to \eqref{eq: rho_c} and using the properties of the projectors  \eqref{eq:proj_identities}, we get our final form for $\rho(\mathbf{w},t)$ as
\begin{equation} \label{eq:solution}
\rho_c(\mathbf{w},t)=\rho_{c0}\Bigl(\exp(-\mathbf{M} t) \cdot \mathbf{w}\Bigr)
\exp\left(-\mathbf{w}^\top \cdot \mathbf{R}_2(t) \cdot \mathbf{w} \right),
\end{equation}  
where
\begin{eqnarray}
 \mathbf{R}_2(t)=\exp(-\mathbf{M} t) \cdot \mathbf{R}_1 (t) \cdot \exp(-\mathbf{M} t)  =  \sum^2_{k,\ell=1}\mathbf{P}^\top_k \cdot \mathbf{R}\cdot \mathbf{P}_\ell 
 \frac{1-\exp\left(-(\Lambda_k+\Lambda_\ell)t\right)}{\Lambda_k+\Lambda_\ell}.
\end{eqnarray}
One can verify by direct substitution that the solution (\ref{eq:solution}) satisfies the master equations \eqref{eq:master_equation} and \eqref{eq:master_eq_rewritten} with the initial condition \eqref{eq:initial_value}.

It is interesting to consider the $t \to \infty $ limit. For finite $\lambda$, the real parts of $\Lambda_{1,2}$ are positive. However, in the range considered, $\omega_d$ is real, correspondingly $\mathrm{Re} \,\Lambda_{3,4}=0$. The argument of the first factor on the right hand side of  equation (\ref{eq:solution}) for large $t$ behaves as
\begin{eqnarray}
\exp(-\mathbf{M} t)\cdot \mathbf{w} &\approx&  \sum^4_{k=3} \exp\left({-\Lambda_k t}\right) \mathbf{P}_k \cdot  \mathbf{w}
\nonumber \\
&=&\begin{pmatrix} 
0 & 0 & 0 & 0 \\
0 & 0 & 0 & 0 \\
0 & 0 & \cos(\omega_d t) & -\displaystyle\frac{2\hbar}{m \omega_d} \sin(\omega_d t) \\
0 & 0 & \displaystyle\frac{m \omega_d}{2 \hbar} \sin(\omega_d t) & \cos(\omega_d t)
\end{pmatrix} \cdot \begin{pmatrix}  \Delta \\ K \\ \delta \\ k \end{pmatrix} \nonumber \\
&=&\begin{pmatrix} 0 \\ 0 \\ \cos(\omega_d t) \delta - \displaystyle\frac{2\hbar}{m \omega_d} \sin(\omega_d t) k\\ \displaystyle\frac{m \omega_d}{2 \hbar} \sin(\omega_d t) \delta + \cos(\omega_d t) k \end{pmatrix}\equiv\begin{pmatrix} 0 \\ 0 \\
s_1(\mathbf{w},t) \\ s_2(\mathbf{w},t)
\end{pmatrix}. 
\end{eqnarray}

Apart from the minus sign, the argument of the second, exponential  factor of (\ref{eq:solution}) in the same large $t \to \infty $ limit is
\begin{eqnarray}
\mathbf{w}^\top \cdot \mathbf{R}_2(\infty) \cdot \mathbf{w}  &=&\mathbf{w}^\top \! \cdot \left(\sum^2_{k,\ell=1}\mathbf{P}^\top_k \cdot \mathbf{R}\cdot \mathbf{P}_\ell 
 \frac{1}{\Lambda_k+\Lambda_\ell} \right) \cdot \mathbf{w} \nonumber\\
 & =&\frac{D_{pp}}{4 \hbar \lambda}\Delta^2 +\frac{\hbar(D_{pp}+8D_{px} \lambda m)}{16 \lambda m^2 \omega_c^2 }K^2.
 \end{eqnarray}
Thus, in the large $t$ limit, the asymptotic characteristic function oscillates as
\begin{eqnarray}
\rho_c(\mathbf{w},t \to \infty) = \rho_{c_0}\Bigl(0,0,s_1(\mathbf{w},t),s_2(\mathbf{w},t) \Bigr) 
\exp\left(- \left(\frac{D_{pp}}{4 \hbar \lambda}\Delta^2 +\frac{\hbar(D_{pp}+8D_{px} \lambda m)}{16 \lambda m^2 \omega_c^2 }K^2 \right)\right), \nonumber \\
\label{eq:asymptotic_in_cm}
\end{eqnarray}

Further stability requirements can be read from (\ref{eq:asymptotic_in_cm}), namely, in the second, Gaussian term, the quadratic coefficients  must be  negative. By eq. (\ref{eq:stability_criterions}), these requirements lead to further stability criteria:
\begin{equation}
D_{pp} \ge 0, \quad \text{and} \quad D_{pp}+8D_{px} \lambda m \ge 0.
\label{eq:eq:stability_criterions_II}
\end{equation}

The asymptotic form of the characteristic function, given in Eq. \eqref{eq:asymptotic_in_cm}, reveals that the long-time behavior of the two-oscillator system differs fundamentally from that of a single oscillator coupled to a thermal bath. In the single-oscillator case, the system asymptotically approaches a unique stationary Gaussian state, as shown in Ref. \cite{Breuer}. In contrast, for the two-oscillator system, a remnant of the initial state persists in the asymptotic regime, encoded in the first term of Eq. \eqref{eq:asymptotic_in_cm}. This contribution is explicitly time-dependent and exhibits undamped oscillations at frequency  $\omega_d$. The absence of damping arises from the specific form of the interaction Hamiltonian \eqref{eq:interaction_term_in_H}, which involves only the CM degrees of freedom, leaving the relative coordinate and momentum unaffected by the system-bath coupling.

We note that this situation can be related to the ultracold gases, where the atoms are trapped harmonically \cite{Stringari1999,Stringari2008,RevModPhysultracold}. In such systems it can arise that three elementary excitations are persistent, and all the others have the same finite damping rate. The three remaining excitation frequencies agree with the three trap frequencies in the three Cartesian directions. This enables the experimental determination of the three trap frequencies. These modes are the so-called Kohn modes \cite{Cornell99,Ketterle99}: their frequencies do not depend on temperature, and they are independent of the quantum statistics, i.e., the frequencies are the same for bosons and for fermions. These properties are based on the Kohn theorem \cite{Kohn61,Reidl2000}, which basically says that the first quantized many-body Hamiltonian for $N$ particles written in terms of Jacobian coordinates has a CM part, which does not couple to the rest (including the interactions). In connection with ultracold cases this theorem is based on the fact that the kinetic energy and the harmonic potential energy are separable in Jacobian coordinates, and the interactions depend only on $(N-1)$ Jacobi coordinates, but not on the first Jacobi coordinate, which is the CM coordinate. The same behavior is observed in our coupled harmonic oscillator system: $x_1-x_2$ is also a Jacobian coordinate, but the bath is coupled to other Jacobian coordinate, the CM coordinate. The part of the Hamiltonian, which depends on $\hat{x},\hat{p}$ in \eqref{Hamiltonian_CM} is harmonic and is uncoupled from the remaining parts. Correspondingly, the oscillations described by the relative operators and having the characteristic frequency $\omega_d$ are persistent: they will not die out at any later times.

In the next subsection, we will give and classify the exact Markovian-coefficients $A,B,C,D$ and  their nontrivial limit cases of the HPZ master equation for the Ohmic spectral density with Lorentz-Drude cutoff function for arbitrary  coupling constant $\gamma$, temperature $T$.

\subsection{Exact Markovian coefficients of the HPZ
master equation with the Lorentz-Drude type Ohmic spectral density}
\label{sub:markovian_coeffs}

The Markovian master equation's coefficients can be obtained from the non-Markovian coefficients $A(t)$, $B(t)$, $C(t)$ and $D(t)$ of Eq.(\ref{eq:HPZ_comm}) in the limit $t \to \infty$, provided these coefficients tend to a constant values. Indeed, this is the case. The actual calculation proceeds  almost parallel to the one for a single oscillator coupled to the bath at zero temperature presented in Ref.~\cite{HBCS}, and this latter work is based on the seminal work of Halliwell and Yu \cite{H-Y}. 

The heart of the problem is the equation of motion of the oscillator belonging to the CM degree of freedom. The effective equation of motion for this oscillator is of the form
\begin{equation}
\frac{d^2}{ds^2}{q}(s)+\Omega^2 q(s)+\frac{2}{M} \int_0^s d \lambda \,\eta(s-\lambda) q(\lambda)=\frac{f(s)}{M},
\label{eq:diff_eqs_for_q(t)}
\end{equation}
where the temperature-independent kernel $\eta(s)$ is fixed by spectral  density $I(\omega)$ of the bath  as
\begin{equation}
\eta(s)=-\int_0^\infty d\omega\, I(\omega)\sin(\omega s).
\label{eq:def_eta}
\end{equation}
The spectral density $I(\omega)$ is the only relevant distributions of  our problem, it is a characteristic of the ensemble of the bath oscillators:
\begin{equation}
I(\omega)=\sum_n \delta(\omega-\omega_n) \frac{C^2_n}{2 m_n \omega_n}.
\label{eq:def_of_spectral_density}
\end{equation}
The form of $f(s)$ in (\ref{eq:diff_eqs_for_q(t)}) , explicitly written out in \cite{H-Y}. 
Eq. (\ref{eq:diff_eqs_for_q(t)}) is a linear, inhomogeneous, integro-differential equation with memory in time, containing all the orders of the interaction, i.e., its validity is not restricted to the weak coupling limit.

The first step is to find the solutions $u_{1,2}$ of the homogeneous equation of motion related to \eqref{eq:diff_eqs_for_q(t)}: 
\begin{equation}
\frac{d^2}{ds^2}{u}(s)+\Omega^2 u(s)+\frac{2}{M} \int_0^s d \lambda \,\eta(s-\lambda) u(\lambda)=0.
\label{eq:diff_eqs_for_u}
\end{equation}
Boundary conditions for $u_1$ and $u_2$ are:
\begin{eqnarray}
u_1(s=0)&=&1, \quad u_1(s=t)=0, \label{eq:bound_cond_for_u1} \\
u_2(s=0)&=&0, \quad u_2(s=t)=1. \label{eq:bound_cond_for_u2}
\end{eqnarray}
Coefficients $A(t)$ and $B(t)$ have been found in \cite{H-Y} to be independent of the temperature and can be expressed as: 
\begin{equation}
A(t)=2\int_0^t ds\, \eta(t-s)u_2(s)-2\frac{\dot{u}_2(t)}{\dot{u}_1(t)} \int_0^t ds\, \eta(t-s)u_1(s),
\label{eq:def_of_a(t)}
\end{equation}
and
\begin{equation}
B(t)=\frac{2}{M\dot{u}_1(t)} \int_0^t ds\, \eta(t-s)u_1(s),
\label{eq:def_of_b(t)}
\end{equation}
respectively.

The coefficients $C(t)$ and $D(t)$ depend on temperature and require also the knowledge of two Green's functions, which are the solutions of the inhomogeneous, linear integro-differential equation belonging to \eqref{eq:diff_eqs_for_q(t)}
\begin{eqnarray}
\frac{d^2}{ds^2}{G_i}(s,\tau)+\Omega^2 G_i(s,\tau)+\frac{2}{M} \int_0^s d \lambda \,\eta(s-\lambda) G_i(\lambda,\tau) 
=\delta(s-\tau), \qquad i=1,2, 
\label{eq:diff_eqs_for_G}
\end{eqnarray}
with boundary conditions for $G_1$ and $G_2$ given  by
\begin{eqnarray}
G_1(s=0,\tau)&=&0, \quad \left.\frac{\partial}{\partial s}G_1(s,\tau)\right|_{s=0}=0, \label{eq:bound_cond_for_G1} \\
G_2(s=t,\tau)&=&0, \quad \left.\frac{\partial}{\partial s}G_2(s,\tau)\right|_{s=t}=0. \label{eq:bound_cond_for_G2}
\end{eqnarray}
Let us adopt the notation  $\dfrac{\partial}{\partial s} G(s, \tau) \equiv G'(s,\tau)$.
Halliwell and Yu determined the form of non-Markovian coefficients $C(t)$ and
$D(t)$ as

\begin{eqnarray}
C(t)&=&\frac{\hbar}{M}\int_0^\infty d\lambda\, G_1(t,\lambda)\nu(t-\lambda) \nonumber \\
&&-\frac{2\hbar}{M^2} \int_0^t ds \int_0^\infty d\tau \int_0^\infty d\lambda \, \eta(t-s)G_1(t,\lambda)G_2(s,\tau)\nu(\tau-\lambda), 
\label{eq:def_of_c(t)} \\
D(t)&=&\hbar\int_0^\infty d\lambda\, G'_1(t,\lambda)\nu(t-\lambda) \nonumber\\
&&-\frac{2\hbar}{M} \int_0^t ds \int_0^\infty d\tau \int_0^\infty d\lambda \, \eta(t-s)G'_1(t,\lambda)G_2(s,\tau)\nu(\tau-\lambda).
\label{eq:def_of_d(t)}
\end{eqnarray}
Here a new temperature dependent kernel shows up:
\begin{equation}
\nu(s)
=\int_0^\infty d\omega\, I(\omega)\coth\left(\frac{\hbar\omega}{2 k_B T}\right)\cos(\omega s).
\label{eq:def_nu}
\end{equation}

To solve the integro-differential equations in Eqs. \eqref{eq:diff_eqs_for_u} and \eqref{eq:diff_eqs_for_G}, one needs a spectral density $I(\omega)$. In this paper, we restrict ourselves to an ohmic spectral density with a Lorentz-Drude type  function \cite{Weiss,book1,HBCSCS}
and a high-frequency cutoff $\Omega_c$:
\begin{equation}
I(\omega)=\frac{2M\gamma \Omega_c^2}{\pi} \frac{\omega}{\omega^2+\Omega_c^2}.
\label{eq:ohmic_spectral_density}
\end{equation}
$\gamma$ is a kind of an effective coupling constant between the bath and the central oscillators. We note that we chose $M$ in Eqs. (\ref{eq:diff_eqs_for_q(t)}) and (\ref{eq:ohmic_spectral_density}) to be the mass of the CM oscillator, i.e. $M=2m$.  

An important special case is the weak coupling limit, which is widely studied in the literature (c.f. \cite{book1} and further references therein). By calculating the coefficients in leading order in the coupling constant $C_n$, Halliwell and Yu also obtained the formulas of a consistent weak coupling limit. 
Now, the time-dependent coefficients read as
\begin{eqnarray}
A_w(t)&=&2\int_0^t ds \, \eta(s) \cos(\Omega s),
\label{eq:a(t)_in_weak_coupling} \\
B_w(t)&=&-\frac{2}{M\Omega}\int_0^t ds \, \eta(s) \sin(\Omega s),
\label{eq:b(t)_in_weak_coupling} \\
C_w(t)&=&\frac{\hbar}{M\Omega}\int_0^t ds \, \nu(s) \sin(\Omega s),
\label{eq:c(t)_in_weak_coupling} \\
D_w(t)&=&\hbar\int_0^t ds \, \nu(s) \cos(\Omega s),
\label{eq:d(t)_in_weak_coupling}
\end{eqnarray}    
where we have indicated this approximation with the index $w$. 

Markovian coefficients are the asymptotic ($t \to \infty$) coefficients $A$, $B$, $C$ and $D$ without arguments.
As the Hamiltonian, belonging to the relative coordinates, completely decouples from the rest in (\ref{Hamiltonian_CM}), and the rest, belonging to the CM degree of freedom, is the same as in Refs. \cite{H-Y,HBCS} for one oscillator coupled by position-position coupling to the bath, correspondingly, one can take the Halliwell-Yu solutions to the master equation's coefficients \eqref{eq:def_of_a(t)}, \eqref{eq:def_of_b(t)}, \eqref{eq:def_of_c(t)}, and \eqref{eq:def_of_d(t)}. 

In case of Lorentz-Drude type spectral density (\ref{eq:ohmic_spectral_density}), for which the intermediate functions $u_1$, $u_2$, $G_1$ and $G_2$ are solved using  Laplace-transforms in \cite{HBCS}, and are best parametrized by the three roots $z_1,z_2,z_3$ of the cubic equation
\begin{equation}\label{eqn:nonshift}
    z^3 + \Omega_c z^2 + \Omega^2 z + \Omega^2 \Omega_c - 2 \gamma \Omega_c^2 = 0.
\end{equation}
Vi\'eta's formulae\footnote{Vieta’s formulae relate the coefficients of an n-th degree algebraic equation to polynomial expressions of its roots (see, for example, \cite{vinberg2003}).} can help to compress some later expressions 
\begin{eqnarray}
z_1 + z_2 + z_3 &=& - \Omega_c, \label{eq:tracetag} \\
\label{eq:spadj}
z_1 z_2 + z_2 z_3 + z_3 z_1 &=& \Omega^2, \\
\label{eq:det}
z_1z_2 z_3 &=& - (\Omega^2 \Omega_c - 2 \gamma \Omega_c^2).
\end{eqnarray}
We order the roots such that $\mathrm{Re}(z_1)$ has the smallest real part.
Stability criteria (\ref{eq:stability_criterions}) 
require that 
\[
\mathrm{Re}\, z_1,\mathrm{Re}\, z_2,\mathrm{Re}\, z_3 <0.
\]
The latter constraints restrict the coupling constant $\gamma$ in the Drude-type spectral density $I(\omega)$ to the range
\[
0\leq \gamma < \gamma_\text{crit}=\frac{\Omega^2}{2\Omega_c}.
\]

\begin{table}
\begin{center}
\begin{tabular}{|c|c|c|c|c|}
\hline
\hline
Physical  Conditions& $  A$ &  $  B$ & $ C$ & $ D$   \\
\hline
\hline
 $T = 0$, & Eq.~\eqref{eq:A(t)_stac} &  Eq.~\eqref{eq:B(t)_stac}  & Eq.~\eqref{eqn:C_exact_zero_temp} & Eq.~\eqref{eqn:D_exact_zero_temp}    \\
\hline
$T = 0$, &  Eq.~\eqref{eq:A_weak}  & Eq.~\eqref{eq:B_weak}   &  Eq.~\eqref{eq:C_w_asymptotic} &    Eq.~\eqref{eq:D_w_asymptotic} \\
weak &&&&\\
\hline
\hline
 $ 0 \leq T \leq \infty$  & Eq.~\eqref{eq:A(t)_stac} & Eq.~\eqref{eq:B(t)_stac}  & Eq.~\eqref{eqn:C_exact} &  Eq.~\eqref{eqn:D_exact}  \\
\hline
$0 \leq T \leq \infty$, &  Eq.~\eqref{eq:A_weak}   &  Eq.~\eqref{eq:B_weak}   & Eq.~\eqref{eqn: C_{2w}}  & Eq.~\eqref{eqn: D2w}       \\
weak &&&&\\
\hline
\hline
 $0 \!< \!\hbar \Omega \ll \hbar \Omega_c \ll k_B T$& Eq.~\eqref{eq:A(t)_stac} & Eq.~\eqref{eq:B(t)_stac}  &Eq.~\eqref{eqn: C3} & Eq.~\eqref{eqn: d3}    \\
\hline
$ 0\! < \!\hbar \Omega \ll \hbar \Omega_c \ll k_B T$, &  Eq.~\eqref{eq:A_weak} &  Eq.~\eqref{eq:B_weak}  & Eq.~\eqref{eqn: C3w} & Eq.\,\eqref{eqn: D3w}      \\
weak &&&&\\
\hline
\hline
\end{tabular}
\end{center}
\caption{Classification of Markovian coefficients' form of the HPZ master equation with Lorentz-Drude type Ohmic
spectral density in different parameter regions for the general  case $0 \leq \gamma \leq \gamma_{cr}$ and for the weak coupling limit $0 \leq \gamma \ll \gamma_{cr}$.}
\label{tab: Table 1} 
\end{table}

Here we show the most general values for the Markovian-coefficients, valid at any temperature.
Several limiting cases are discussed in Appendix~\ref{sec:markovian_coeffs}.
The temperature independent coefficients $A$ and $B$ are:
\begin{equation}
A=-M z_1 (z_2+z_3), \qquad B=-z_2-z_3.
\label{eq:A_and_B}
\end{equation}

The temperature dependent coefficients $C$ and $D$ are:
\begin{eqnarray}
C&=& \frac{k_B T(z_2+z_3)}{z_1}+\frac{\hbar(z_2+z_3)(z_1^2+z_2 z_3)}{\pi(z_1-z_2)(z_1-z_3)}K(z_1,\nu) \nonumber \\
&&+ \frac{\hbar(z_1+z_3)(z_2+z_3)z_2}{\pi (z_1-z_2)(z_3-z_2)}K(z_2,\nu)
+\frac{\hbar(z_1+z_2)(z_2+z_3)z_3}{\pi (z_1-z_3)(z_2-z_3)}K(z_3,\nu),
\label{eqn:C_exact}
\end{eqnarray}
and
\begin{eqnarray}
D&=& -k_B T M (z_2+z_3)+\frac{\hbar M z_1^2(z_2+z_3)^2}{\pi(z_1-z_2)(z_1-z_3)}K(z_1,\nu) \nonumber \\
&&+\frac{\hbar M(z_1+z_3)(z_2+z_3)z_2^2}{\pi (z_1-z_2)(z_3-z_2)}K(z_2,\nu)
+ \frac{\hbar M(z_1+z_2)(z_2+z_3)z_3^2}{\pi (z_1-z_3)(z_2-z_3)}K(z_3,\nu), \label{eqn:D_exact}
\end{eqnarray}
where the function $K(z,\nu)$ is defined in Eq.~(\ref{eq:def_of_K(z,nu)}).

For completeness, we provide Table.~\ref{tab: Table 1}, indicating several limiting cases together with the relevant equation numbers in Appendix~\ref{sec:markovian_coeffs}.

Finally, it is important to note that exact formulas for computing the coefficients are known (see Refs. \cite{HPZ, H-Y}). However, their evaluation involves solving both homogeneous and inhomogeneous integro-differential equations (see Eqs. \eqref{eq:diff_eqs_for_u} and \eqref{eq:diff_eqs_for_G}), which are generally challenging to deal with. In contrast, expressions in the weak coupling limit are much more tractable and are already provided in Refs. \cite{HPZ, H-Y}. In Ref. \cite{Ford}, explicit expressions for the coefficients were derived, but we were only able to confirm agreement for $A(t)$ and $B(t)$. We believe the discrepancies in the coefficients $C(t)$ and $D(t)$ result from the differences in methodology. Specifically, we have solved the homogeneous and inhomogeneous integro-differential equations exactly, whereas in Ref. \cite{Ford} the memory term was handled by a time-local approximation.

\section{Quantum mutual information, entanglement and  logarithmic negativity  in case of Gaussian density operators}\label{sec:III}
	
In this section, we shall use units such that $\hbar=1$.

The best representation for the solution of the master equation was the $\rho_c$ characteristic function \eqref{rho_in_cm} in CMR phase-space variables \eqref{eq:CM_vect}, due to the special form of the coupling to the bath oscillators in the microscopic Hamiltonian \eqref{Hamiltonian_CM}. However, entanglement and related properties are bounded to the original operators $\hat{\mathbf{r}}^\top=(\hat{x}_1,\hat{p}_1,\hat{x}_2,\hat{p}_2)$ and variables $\mathbf{r}^\top=(\Delta_1,k_1, \Delta_2,k_2)$.
The two sets of operators and phase-space variables are connected by linear transformations
\begin{equation}
 \hat{\mathbf{r}}=\mathbf{W}\hat{\mathbf{w}},\qquad \mathbf{r}=\mathbf{W}\mathbf{w}, 
\label{eq:lin_coord_traf}
\end{equation}
where
\begin{equation}
\mathbf{W}= \begin{pmatrix} 1&0&\frac{1}{2}&0\\ \noalign{\medskip}0&\frac{1}{2}&0&1
\\ \noalign{\medskip}1&0&-\frac{1}{2}&0\\ \noalign{\medskip}0&\frac{1}{2}&0&-1
\end{pmatrix}.
\end{equation}
$\mathbf{W}$ is a symplectic matrix \cite{Serafini}: $\mathbf{W}  \in \textrm{Sp}_{4,\mathbb{R}}$.

 If the initial characteristic function $\rho_{c_0}(\mathbf{w})$ in phase-space variables $\mathbf{w}$ is Gaussian, it remains in the Gaussian time evolution, i.e., it is Gaussian for all the times $t$ \cite{Fleming}.  
Gaussian characteristic functions in variables $\mathbf{r}$ have the form \cite{Serafini}
\begin{equation}
\rho^{(G)}(\mathbf{r},t)=\exp{\left( -\frac{1}{4} \mathbf{r}^\top \Omega^\top \pmb{\sigma}(t) \Omega \mathbf{r}+i \mathbf{r}^\top \Omega^\top \overline{\mathbf{r}}(t)\right)},
\label{eq:gaussian_char_func}
\end{equation}
where the skew-symmetric matrix $\boldsymbol{\Omega}$ is
\begin{equation}
 \boldsymbol\Omega = \begin{pmatrix}
	 \boldsymbol\Omega_1 & \mathbf{0} \\
	 \mathbf{0} &  \boldsymbol\Omega_1
\end{pmatrix}
 , \quad \text{and} \quad 
  \boldsymbol\Omega_1 = \begin{pmatrix}
	0 &  1 \\
	-  1 & 0
\end{pmatrix}, 
\end{equation}
with  (possibly) time-dependent first moments
\begin{equation}
\overline{\mathbf{r}}_i(t)=\mathrm{Tr}\left[ \hat{\rho}^{(G)}(t) \hat{\mathbf{r}}_i\right],\qquad i=1,\ldots,4,
\end{equation}
and the (possibly) time-dependent Gaussian covariance matrix $\pmb{\sigma}(t)$ 
\begin{equation}
\pmb\sigma_{i,j}(t)=\mathrm{Tr}\left[ \hat{\rho}^{(G)}(t) \left\{\hat{\mathbf{r}}_i-\overline{\mathbf{r}}_i(t),\hat{\mathbf{r}}_j-\overline{\mathbf{r}}_j(t)\right\}\right].
\end{equation}
In the last equation, $\{ ,  \}$ denotes an anti-commutator. By definition
$\pmb\sigma_{i,j}(t)$ is a symmetric and positive definite matrix.

As a consequence of the linear transformations \eqref{eq:lin_coord_traf}, a similar form to \eqref{eq:gaussian_char_func} is true for the corresponding Gaussian operator $\hat{\rho}_c$ in variables $\mathbf{w}$, but with a different  first moment $\overline{\mathbf{w}}_c(t)$ and covariance matrix $\pmb{\sigma}_c(t)$. Trivially they are connected by the transformations:
\begin{equation}
\overline{\mathbf{r}}(t)=\mathbf{W} \overline{\mathbf{w}}_c(t), \qquad \pmb{\sigma}(t) =\mathbf{W}\pmb{\sigma}_c(t) \mathbf{W}^\top.
\label{eq:transformation_of_momenta}
\end{equation}
Taking a certain initial $\pmb{\sigma}(t=0)$ matrix and $\overline{\mathbf{r}}(t=0)$ vector, \eqref{eq:solution} predicts $\pmb{\sigma}(t)$ and $\overline{\mathbf{r}}(t)$ at any later times for the Markovian master equation. These time dependence shall we use in the following, especially in our numerical examples.

The very special properties of Gaussian kernels as integral operators are that \cite{Serafini} their eigenvalues are products of geometric series, and the quotients of the geometric series are directly related to the symplectic eigenvalues of the covariance matrix. More precisely, the eigenvalues of the Gaussian operators (density matrices) for two particles  
are given by
\begin{eqnarray}
		\lambda_{N_1,N_2}&=&\frac{4}{(\nu_1+1)(\nu_2+1)} \left( \frac{\nu_1-1}{\nu_1+1}\right)^{N_1} \left( \frac{\nu_2-1}{\nu_2+1}\right)^{N_2}, 
        \qquad  N_1,N_2=0,1,\ldots,. \label{eq:Neumann_ent_gen_tot}
\end{eqnarray}
where the $\nu_1$ and $\nu_2$ symplectic eigenvalues are the positive eigenvalues of the matrix $i \boldsymbol{\Omega}\cdot\pmb{\sigma}(t)$ (the remaining two eigenvalues are $-\nu_1$ and $-\nu_2$). Of course, for time-dependent covariance matrix $\pmb\sigma(t)$, the Gaussian eigenvalues $\lambda_{N_1,N_2}$, and the symplectic eigenvalues $\nu_1$ and $\nu_2$ also depend on time, but for simplicity, we do not write out the explicit time-dependence. 

From Eq.~\eqref{eq:Neumann_ent_gen_tot} is clear, that positivity of the spectrum requires
\begin{equation}
\nu_1>1, \qquad \nu_2>1.
\label{eq:positivity_of_gauss_spectr}
\end{equation}

The covariance matrix can be partitioned to two by two blocks as follows
\begin{equation}
\pmb{\sigma}=\begin{pmatrix}
\pmb{\sigma}_{11} & \pmb{\sigma}_{12} \\
\pmb{\sigma}_{21} & \pmb{\sigma}_{22}
\end{pmatrix}
\label{eq:partitioning_of_total_sigma}
\end{equation}
in that phase space where $\mathbf{r}$ is used.

The two central oscillators that are coupled to the bath can be considered as subsystems $A$ and $B$. The individual characteristics of the two subsystems can be obtained in the usual way: by taking partial trace in the density operator of the total system $A+B$. If the total system is Gaussian, the biggest advantage of the characteristic function is that the density operators of the subsytems
\begin{equation} \label{eq:Neumann_ent_sub_gen}
\hat{\rho}_A=\mathrm{Tr}_{B} \hat{\rho}, \quad  \hat{\rho}_B=\mathrm{Tr}_{A} \hat{\rho},
\end{equation}
are also Gaussian with covariance matrices
\begin{equation}
\pmb{\sigma}_A=\pmb{\sigma}_{11}, \qquad \pmb{\sigma}_B=\pmb{\sigma}_{22},
\end{equation}
see Eq.~\eqref{eq:partitioning_of_total_sigma}.
Let us denote the symplectic eigenvalues of $\pmb{\sigma}_A$ and $\pmb{\sigma}_B$ by $\nu_A$ and $\nu_B$, respectively, i.e., $\nu_A$ and $\nu_B$ are the positive eigenvalues of $i\pmb{\Omega}_1\pmb{\sigma}_A$ and $i\pmb{\Omega}_1\pmb{\sigma}_B$. We also note that since $\mathbf{W}$ is a symplectic matrix, therefore the symplectic eigenvalues for ${\pmb{\sigma}}_c$ and ${\pmb{\sigma}}$ are the same.

One of the most important concept of quantum information theory is the von Neumann entropy of the total system $A+B$, which is defined as \cite{Petz2007}
\begin{equation} \label{eq:Neumann_ent_Gauusian_tot}
		S_{A+B}=-\mathrm{Tr} \hat{\rho} \ln \hat{\rho} .
\end{equation}
For the Gaussian spectrum \eqref{eq:Neumann_ent_gen_tot} it is 
\begin{eqnarray}
		S_{A+B}&=&
        - \sum^{\infty}_{N_1=0} \sum^{\infty}_{N_2=0} \lambda_{N_1,N_2} \ln \left( \lambda_{N_1,N_2}\right)= \nonumber\\
     &=&  \frac12 \left( 1+\nu _{1} \right) \ln  \left( 1+\nu _{1} \right) +
		\frac12 \left( 1+\nu _{2} \right) \ln  \left( 1+\nu _{2} \right)  \nonumber \\
		&&+\frac12\left(1-\nu_{1} \right) \ln  \left( \nu _{1}-1 \right) 
		+\frac12\left( 1-\nu _{2} \right) \ln  \left( \nu _{2}-1
		\right) -2\,\ln{2}.  \label{eq:etropy_for_2Gauss}
\end{eqnarray}
The reduced density operators of subsystems $A$ or $B$ are defined in \eqref{eq:Neumann_ent_sub_gen}, thus, the von Neumann entropies of subsystems reads as:
\begin{equation}
S_{X}=-\mathrm{Tr} \hat{\rho}_X\ln{\hat{\rho}_X}, \qquad  X \in \{ A,B \}.
\end{equation}
It is easy to check that
\begin{eqnarray} 
S_{X}&=& -\ln   2+\frac12 \left( 1+\nu_{X} \right) \ln  \left( 1+\nu _{X} \right) \nonumber \\
&&+	\frac12\left(1-\nu_{X} \right) \ln  \left( \nu_{X}-1 \right) 
			, \qquad X \in \{A,B\}.  \label{eq:Neumann_ent_sub_Gaussian}
\end{eqnarray}
The Peres–Horodecki criterion \cite{Perescrit,HORODECKI19961} has been extended to continuous variable systems by Simon \cite{SimonR}, which uses the so called partial transpose transformation. Let as choose the reflection (partial transpose) of $k_2 \to -k_2$ described by the matrix 
\begin{equation}
\pmb{\tau}=\begin {pmatrix} 1&0&0&0\\ \noalign{\medskip}0&1&0&0
\\ \noalign{\medskip}0&0&1&0\\ \noalign{\medskip}0&0&0&-1\end {pmatrix}.
\end{equation}
The covariance matrix after partial transpose is
\begin{equation} \label{eq:par_transp}
\pmb{\sigma}^{PT}=\pmb{\tau}\cdot \pmb{\sigma} \cdot \pmb{\tau}
\end{equation}
with symplectic eigenvalues\footnote{  To avoid possible misunderstandings, we emphasize that the superscript $^{*}$ should not be confused with the usual notation for complex conjugation; it is merely our convention to denote the symplectic eigenvalues after partial transposition, which are real.} $\pmb{\nu}^{_*}$. If any of the elements of $\pmb{\nu}^{_*}$ are smaller than one, then the Gaussian state associated with $\pmb{\sigma}$ is entangled.

The Schatten -- von Neumann norm  of a self-adjoint  trace class operator $\hat{O}=\hat{O}^\dagger$ \cite{Schatten2021}

\[
\left\| \hat{O} \right\|_1= \textrm{Tr}\,\sqrt{\hat{O}^\dagger \hat{O}}.
\]
Of course, $|| \hat{O} ||_1=\textrm{Tr}\,\sqrt{\hat{O}^2}=\textrm{Tr}\,|\hat{O}|<\infty$.  If  $|\textrm{Tr}\,\hat{O}|< ||\hat{O}||_1$, then $\hat{O}$ has negative eigenvalues. For  Gaussian kernels, it
is reducing to 

\begin{eqnarray}
		\|\hat{O}^{(G)} \|_1&=&\sum^{\infty}_{N_1=0}\sum^{\infty}_{N_2=0}|\lambda_{N_1,N_2}| 
         = \frac{4}{\left(|\nu_1-1|-(1+\nu_1)\right) \left(|\nu_2-1|-(1+\nu_2)\right)}.\phantom{MM} \label{eq:Schatten}
\end{eqnarray} 
 We will use this fact for checking entanglements by the Peres-Horodecki criterion.
A related quantity, called logarithmic negativity is \cite{Logneg2}:
\begin{equation}  \label{eq:log_neg}
		E_N\left( \hat{\rho} \right)= \log_2 \|\hat{\rho}^{(PT)} \|_1=\frac{ \ln{ \|\hat{\rho}^{(PT)} \|_1}}{\ln{2}}.
\end{equation}
It is also a measure of entanglements, a full entanglement monotone, however it is not convex \cite{Logneg2}. It is important to emphasize that, in our case, the Werner-Wolf theorem \cite{WW} guarantees that when $E_N\left( \hat{\rho} \right)=0$, the Gaussian quantum state of the total system $A+B$ is separable.

Another well-known measure of quantum correlations is the quantum mutual information \cite{nielsen_chuang_2010}. Its definition is inspired by the classical counterpart, which measures the amount of information that two random variables provide about each other \cite{Shannon48}. The quantum mutual information can also be expressed through the quantum relative entropy \cite{umegaki,Petz2007, Correlation_proof}. For the case of two oscillators, it takes the following form:
\begin{equation} \label{eq:mutual_bip}
C_{\xi}=S_A+S_B-S_{A+B} \geq 0.
\end{equation}
We keep the index $\xi$ to make a clear distinction from the coefficient $C(t)$.

Quantum mutual information $C_\xi$ has the following properties: i) it is a "correlation measure", measuring how far the state is from a product state, ii) it also makes sense for mixed states, iii) and it is a good quantity both classically and quantum mechanically, iv) it can be non-zero in a pure state \cite{Szalay_B_app,Szalay_ent_meas}. In our numerical work, we will study  its behavior as a function of time especially, how it behaves for our two oscillator problem.

We note that we have developed more tools, i.e., analytical results, which can be very useful in practice, and are of great help in the thorough investigation of non-Gaussian time evolution, too \cite{Newton,japcsi,approx_traceclass}.

\section{Numerical Results} \label{sec:IV}

\subsection{Initial condition and physical parameters used in numerical calculations}

In the Markovian case, where all the coefficients in the master-equation \eqref{eq:HPZ_comm} are constant, we have the full, time-dependent solution \eqref{eq:solution} of the characteristic function, therefore in principle, we can choose any initial $\rho_c(\mathbf{w},t=0)$ or $\rho(\mathbf{r},t=0)$. A little problem arises at that point to check whether this initial state belongs to a physical density operator. To ensure this, we choose the initial state to be a Gaussian, where several criteria \cite{Serafini} exist to check it. Even we restrict ourselves to zero first momenta $\overline{\mathbf{w}}=0$ or $\overline{\mathbf{r}}=0$.

As an initial covariance matrix, we choose a symmetric matrix used for Einstein-Podolsky-Rosen-state extended with a multiplicative parameter $p\geq 1$:
\begin{eqnarray}
&&\pmb{\sigma}_c(0)=p \cdot \pmb{\sigma}_{\text{EPR}} \nonumber \\ 
&&=p \begin {pmatrix} \cosh \left( 2r_s \right) &0&\sinh
 \left( 2r_s \right) &0\\ \noalign{\medskip}0&\cosh \left( 2r_s
 \right) &0&-\sinh \left( 2r_s \right) \\ \noalign{\medskip}\sinh
 \left( 2r_s \right) &0&\cosh \left( 2r_s \right) &0
\\ \noalign{\medskip}0&-\sinh \left( 2r_s \right) &0&\cosh \left( 2r_s
 \right) \end {pmatrix}. \nonumber \\
\end{eqnarray}
For $p=1$ this is the covariance matrix of a single Einstein-Podolsky-Rosen-state with an EPR-squeezing amplitude $r_s$ \cite{Serafini}.
We note that this covariance matrix belongs to the CMR phase-space description and not where the coordinates of system A and B are defined. The two covariance matrix are still connected  by the transformation in Eq.~\eqref{eq:transformation_of_momenta}.

In a general situation, a Gaussian state
of any number of modes is separable across a bipartition of the modes $A,B$ if and only if its the Gaussian covariance matrix $\pmb{\sigma}$ satisfies $\pmb{\sigma} \geq \pmb{\sigma}_A \oplus \pmb{\sigma}_B$, where $\pmb{\sigma}_A$ and $\pmb{\sigma}_B$ are the covariance matrices of the subsystems $A$ and $B$ \cite{Serafini}. By direct calculation,  $\rho(0)$ is never separable for any $r_s$ and $p$.

The dimensionless external physical parameters, that we do not change during numerical calculations, are summarized in the table (\ref{tab_parameters}).
\begin{table}[ht!]
    \centering
    \begin{tabular}{|c|c|}
    \hline
       $\Omega_c/\Omega$ & $ \gamma / \Omega$   \\ \hline
     40   &  1/128  \\
   \hline
    \end{tabular}
    $\implies$
    \begin{tabular}{|c|c|c|}
    \hline
       $z_1/\Omega$  & $z_2/\Omega$ & $ z_3/\Omega$   \\ \hline
   $-39,9844$ &    $-0,007814 + 0,6124 i$  &  $-0,007814 - 0,6124 i$ \\
   \hline
    \end{tabular}
    \caption{The dimensionless fixed parameters used in the numerics and the numbered roots of the cubic equation \eqref{eqn:nonshift} in our numerical examples.}
    \label{tab_parameters}
\end{table}

Other parameters that will be  varied during our numerical  study are: mixing parameter $p$, squeezing parameter $r_s$, dimensionless temperature $k_B T/(\hbar\Omega)$, coupling constant between the two original oscillators $\kappa$ and the dimensionless time parameter $\Omega t$.

\subsection{Positivity violations in some time evolutions}

As a preliminary remark we note that for one oscillator, it was enough to check the purity from the point of view of the positivity of the Gaussian state \cite{HBCSCS,HBCS}: if $\text{Tr} \hat{\rho}^2(t)>1 $ at any time the Gaussian state is not physical. 

For two oscillators, which is our case, 
$0<\text{Tr} \hat{\rho}^2(t) \leq 1$ is only a necessary condition for positivity, but not a sufficient one. It is expressed in terms of the two
symplectic eigenvalues and 
\begin{equation}
\text{Tr} \hat{\rho}^2(t)=\frac{1}{\nu_1(t)\nu_2(t)} \sim \frac{1}{p^2}.
\end{equation}
By our numerical experience, it is monotonically decreasing as a function of the temperature $T$. Here we do not plot the time evolution of purity, but we checked it. At first glance, it may seem surprising that the analytical expressions for $\nu_1(t)$ and $\nu_2(t)$ are independent of the coupling parameter $\kappa$. However, this is a direct consequence of the fact that the $\kappa$-dependent unitary evolution of the relative operators in Eq. \eqref{eq:HPZ_comm} preserves the eigenvalues of the joint density operator expressed in terms of the relative and center-of-mass degrees of freedom. Likewise, the symplectic transformation in Eq. \eqref{eq:lin_coord_traf}, which reconstructs the positions and momenta of the two oscillators, also leaves the eigenvalues unchanged. Oscillations involving the $\kappa$-dependent frequency
$\omega_d$ (see Eq. \eqref{eq: new_frequencies}) appear only in quantities that involve operations performed after the partial trace or involve the partial transpose.

In the numerics, we have chosen such parameters discussed in the previous subsection that fulfilled stability criteria \eqref{eq:stability_criterions},\eqref{eq:eq:stability_criterions_II} of the asymptotic state of the master equation's solution.
We experienced the following behaviors: i)
The smallest symplectic eigenvalue before partial transpose $\nu_1$  monotonically decreases as a function of the squeezing parameter $r_s$. 
ii) The largest  symplectic eigenvalue before partial transpose $\nu_2$ monotonically grows as a function of the squeezing parameter $r_s$. 
iii) Positivity violation can occur as a function of time at low temperature $T$, for large squeezing parameter $r_s$, or small mixing parameter $p$.

\begin{figure}
\begin{center}
\includegraphics[width=0.6\linewidth, angle=270]{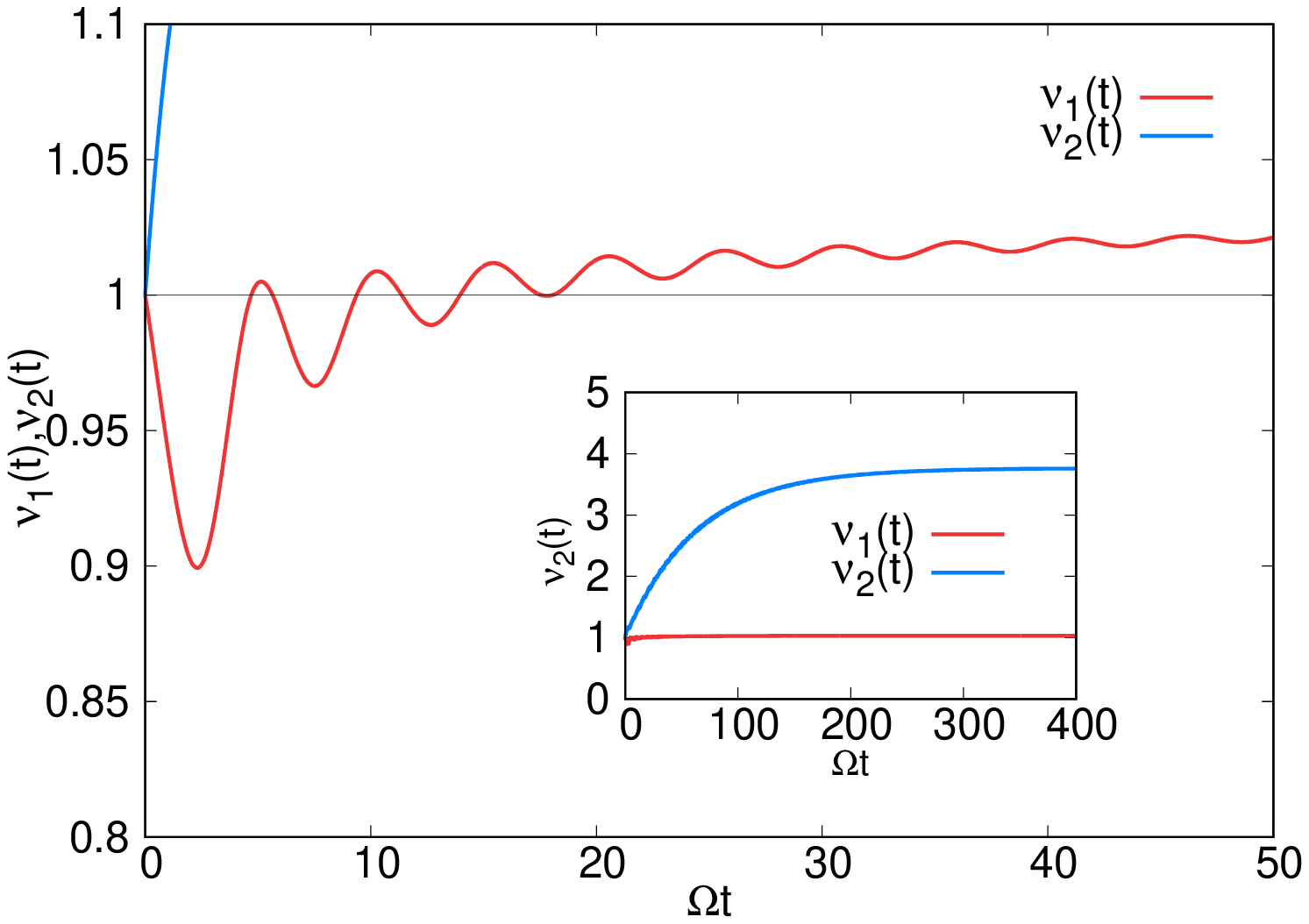}
\end{center}
\caption{ 
Positivity violation ($\nu_i(t)<1$) in Gaussian time evolution shown by the  symplectic eigenvalues. As parameters of the initial covariance matrix and parameters of the master-equation we have chosen $r_s=1$, $p=1$, $k_B T/(\hbar\Omega)=0.1$, $\kappa=-0.2\, m\Omega^2$. Limiting value of $\nu_1(t)$ is: $\nu_1(\infty)=1.03092$.  All quantities plotted here are taken from the analytical solution of the Markovian HPZ master equation (see Eq.~\eqref{eq:solution}). }
\label{fig:pos_viol}
\end{figure}

In Fig.~\ref{fig:pos_viol}, we demonstrate that even with exact Markovian coefficients,  positivity violation at short-times can occur during the time evolution for certain initial conditions. The short-term behavior of the smaller symplectic eigenvalue shows that it goes below $1$ at certain time intervals, which implies that in such cases the solution to the master equation is not physical. It is now clear, why we have extended the EPR covariance matrix with the new parameter $p$: we wanted to avoid prompt positivity violations as seen in Fig.~\ref{fig:pos_viol} by a possible increase of this parameter.

\subsection{Time evolution of the total and the partial von Neumann-entropies, and the quantum mutual information}

\begin{figure}
\begin{center}
\includegraphics[width=0.6\linewidth, angle=270]{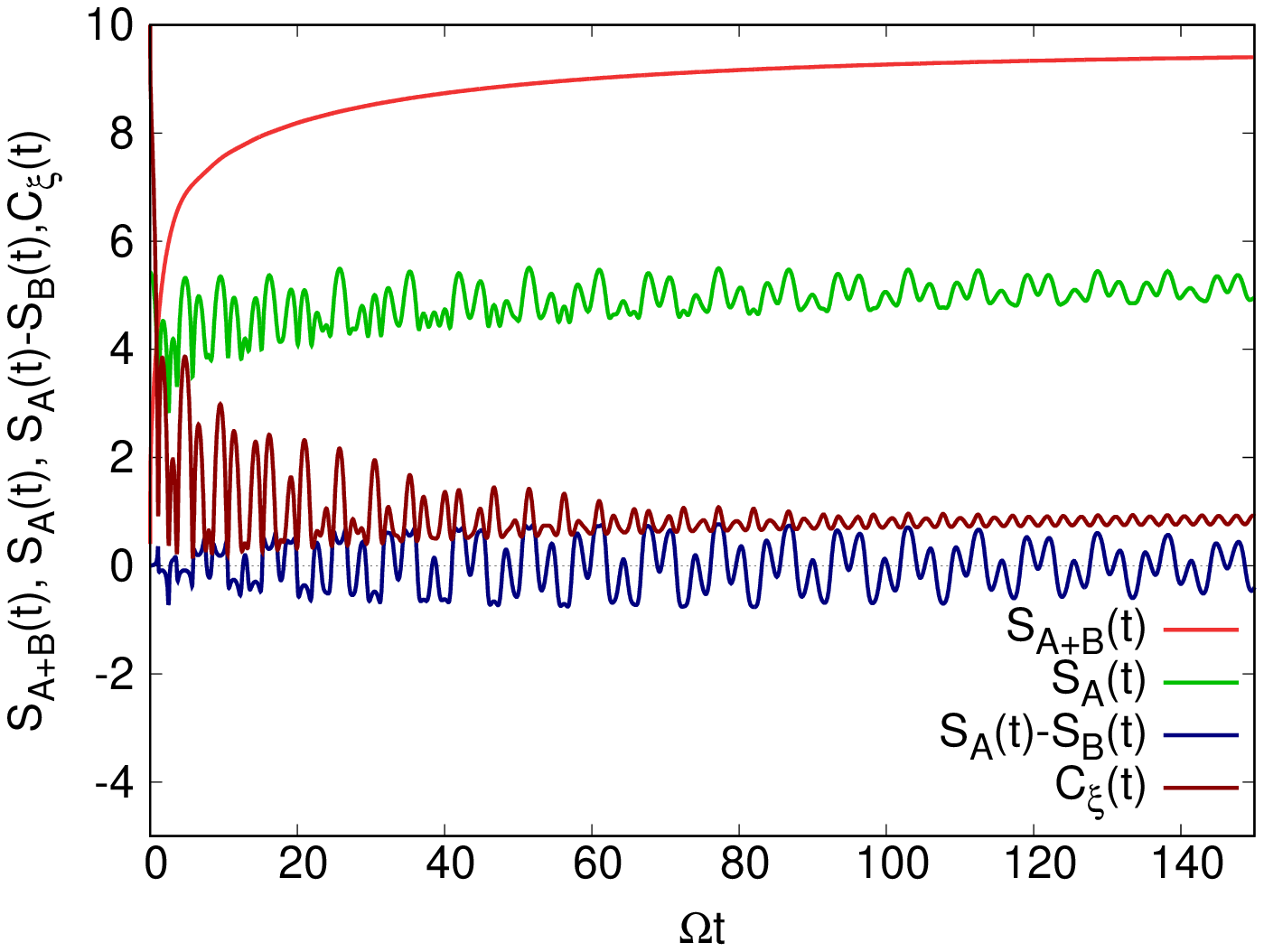}
\end{center}
\caption{ 
The entropy $S_{A+B}(t)$ of the compound system $A+B$ ($A$ and $B$ correspond to the first and second oscillator), the entropy $S_A(t)$, the entropy difference $S_A(t)-S_B(t)$ and the correlation $C_\xi(t)$ as a function of time. Initial state: modified EPR state with $r_s=3$, and with amplitude $p=1.1$. Other parameters are: $k_B T/(\hbar\Omega)=10$, $\kappa=0.2\, m \Omega^2$).   All quantities plotted here are taken from the analytical solution of the Markovian HPZ master equation (see Eq.~\eqref{eq:solution}).
\label{fig:entropies}}
\end{figure}

We have already mentioned that the characteristic equation for the symplectic eigenvalues is independent of the frequency $\omega_d$, correspondingly the von Neumann-entropy \eqref{eq:etropy_for_2Gauss} will not show oscillations related to $\omega_d$. This is clearly seen in Fig.~\ref{fig:entropies}, where the partial entropies $S_A$ and $S_B$ have some additional oscillations. Indeed, analytic calculations show that they depend on $\omega_d$. The same is true for the mutual information \eqref{eq:mutual_bip}: The $\omega_d$ dependence still remains in the sum $S_A+S_B$.  The oscillations are more regular after the transients died out.

The von Neumann-entropy is monotonically increases as a function of time. This behavior is well known for Lindblad master-equation \cite{Kobayashi}. Our master equation is not of Lindblad-form, however,  the von Neumann-entropy still grows as time goes on. This behavior is plausible, but we do not have a proof yet.

Fig. (\ref{fig:entropies}) nicely illustrates our general numerical experience of how the effects of decoherence manifest themselves in the physical, non-unitary  time evolution of entropy related quantities shown.

\subsection{Numerical study of entanglement}

 Here we will study how decoherence affects mutual information and quantum entanglement. We raised the question for our Markovian dynamics: do the Gaussian time evolutions of logarithmic negativity (see \eqref{eq:Schatten} and \eqref{eq:log_neg}) and mutual information, which have completely different physical meanings, show similarities or they are unrelated?

\begin{figure}
\begin{center}
\includegraphics[height=0.78\linewidth, angle=270]{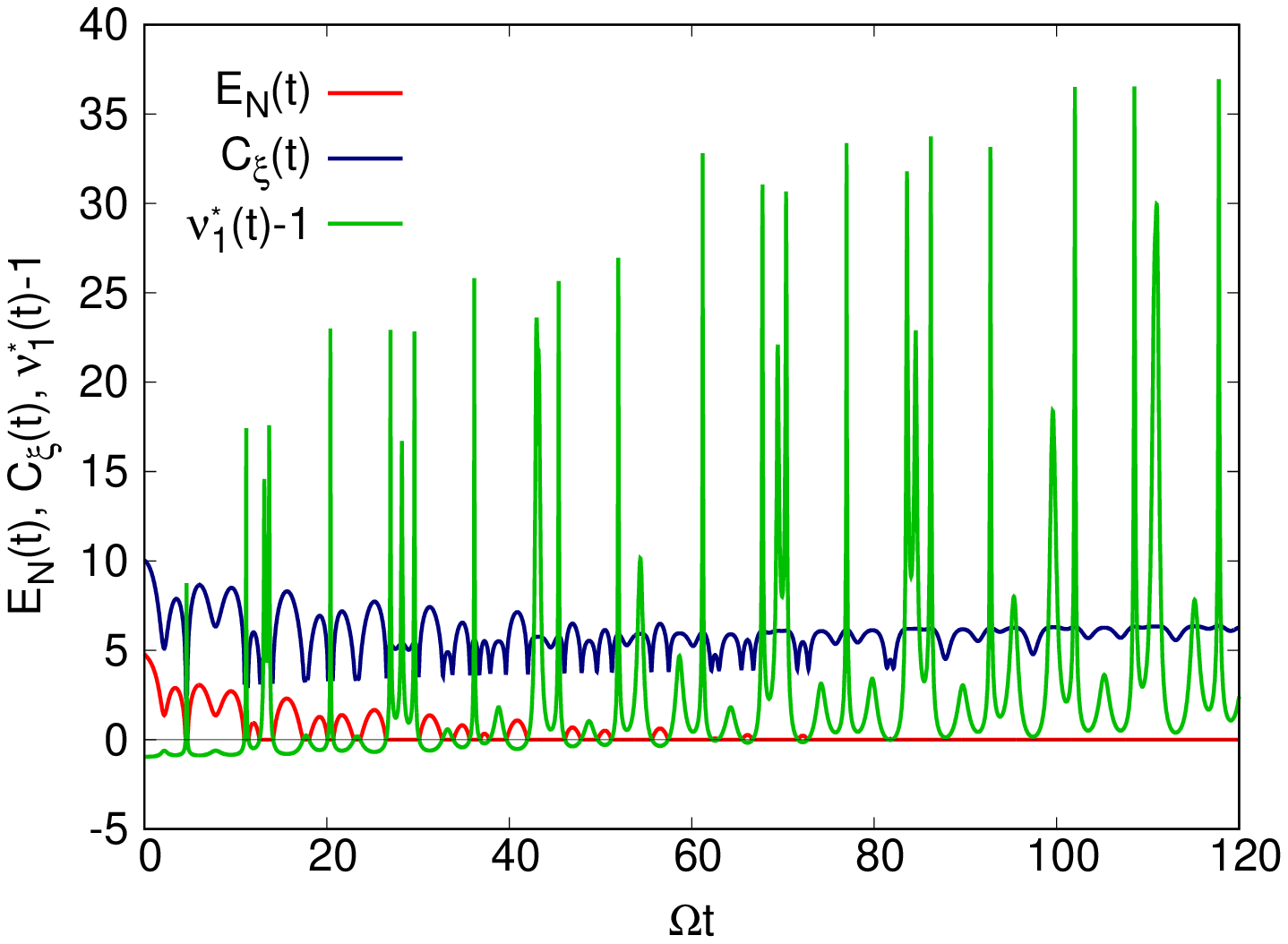}
\qquad
\includegraphics[height=0.78\linewidth, angle=270]{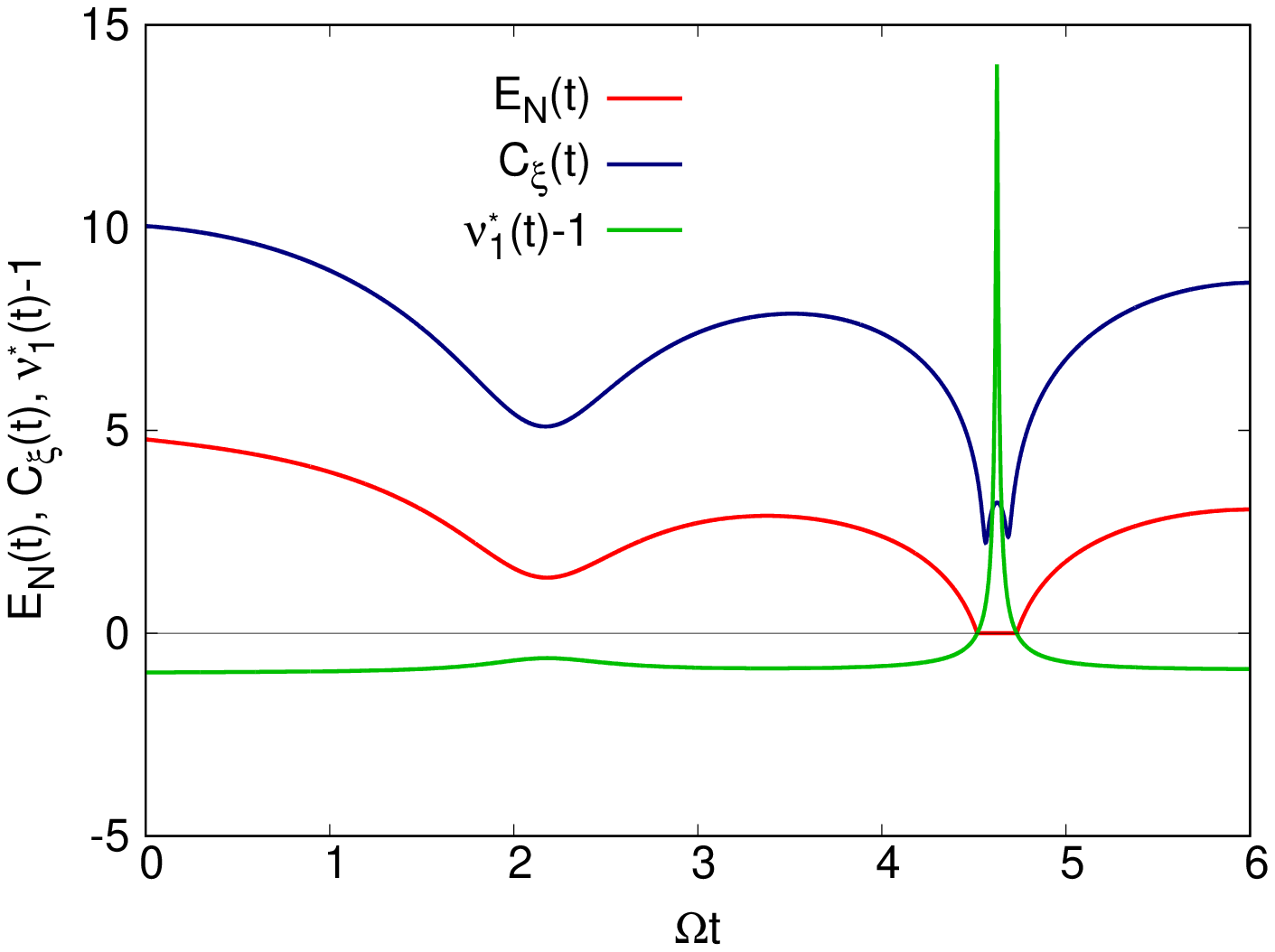}
\end{center}
\caption{ 
Short time behavior of logarithmic negativity $E_N$,  quantum mutual information $C_\xi$, and after partial transpose the smaller symplectic eigenvalue $\nu_1^{_*}$ minus one. For separable states $E_N=0$, $\nu_1^*-1 \geq 0$.
Other parameters are $p=11$, $r_s=3$, $k_B T/(\hbar\Omega)=0.0001$, the coupling constant is $\kappa=0$. The bottom subfigure is a magnification of the top subfigure.
For better comparison, we draw the zero level with a black line. All quantities plotted here are taken from the analytical solution of the Markovian HPZ master equation (see Eq.~\eqref{eq:solution}).}
\label{fig:entangl_kappa_0}
\end{figure}

In Fig.~\ref{fig:entangl_kappa_0}, we show a very interesting phenomenon, consistent with the nonlocality of quantum physics, when we choose the coupling parameter $\kappa$ between the two oscillators to be zero. Despite the zero direct coupling, the total open system still becomes entangled and then disentangled as a function of time due to the bath. In this situation after $\Omega t_s\approx 82$ the quantities logarithmic negativity $E_N(t)$  and $\nu_1^*(t)-1$, which are most relevant from the point of view of the entanglement, predict separable behavior: $E_N(t)$ remains zero, $\nu_1^*(t)-1$ is non-negative. By our numerical experience the characteristic time $t_s$ after which the states are separable, decreases with the increase of the temperature $T$ or the increase of the mixing parameter $p$.

More important feature can be seen on right lower subfigure of Fig.~\ref{fig:entangl_kappa_0}: $E_N(t)$ and $\nu_1(t)-1$ clearly mark the separable situations and the quantum mutual information also mimic the behavior of $E_N(t)$: it is the smallest at the borders when the separable-entangled transition occurs. Of course, not perfectly, from $C_\xi(t)$, one cannot determine whether the system is separable or not, but it's consistent with the meaning that the quantum mutual information is smaller for a less correlated state.    

\begin{figure}
\begin{center}
\includegraphics[height=0.75\linewidth, angle=270]{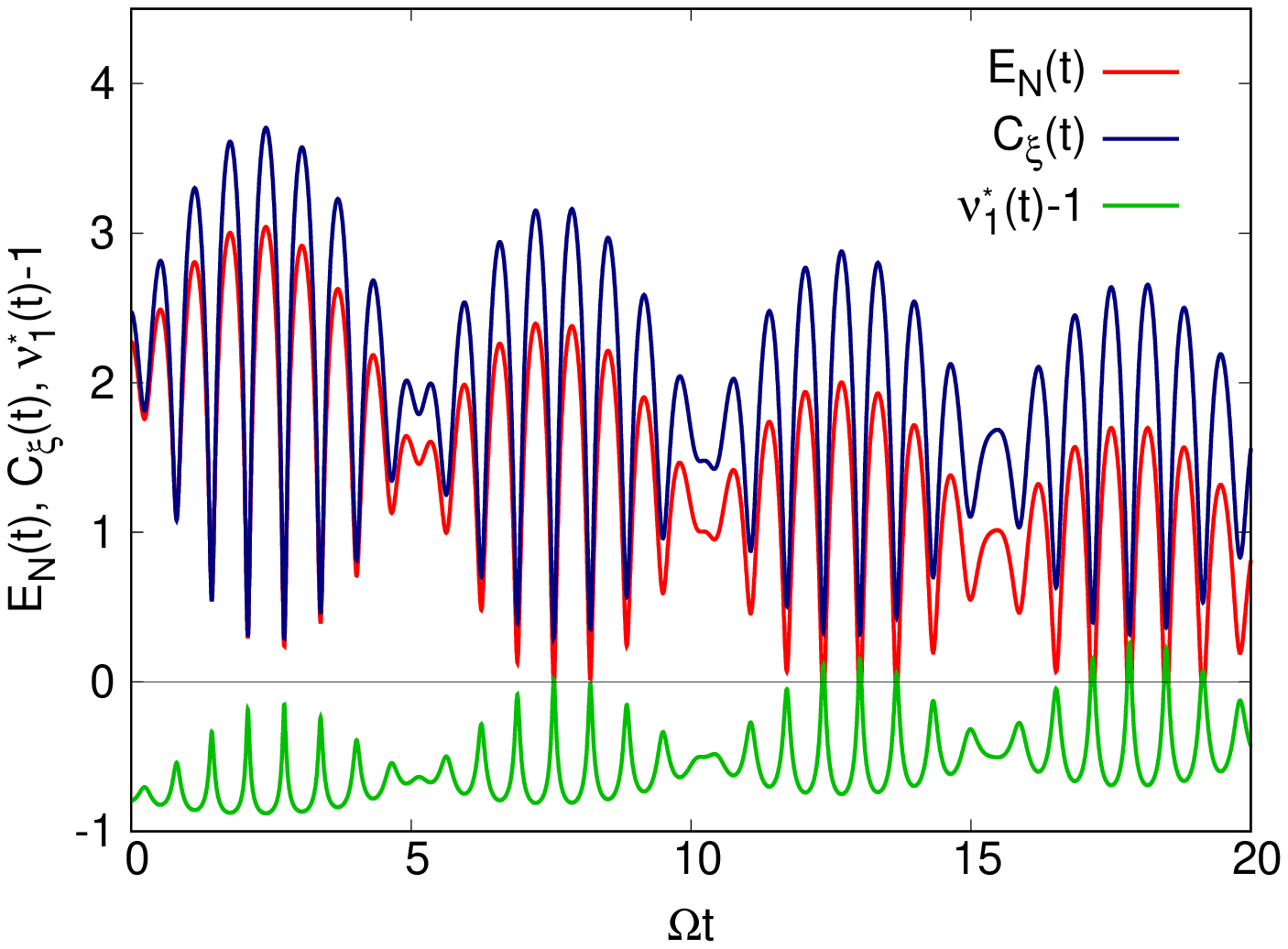} \qquad
\includegraphics[height=0.75\linewidth, angle=270]{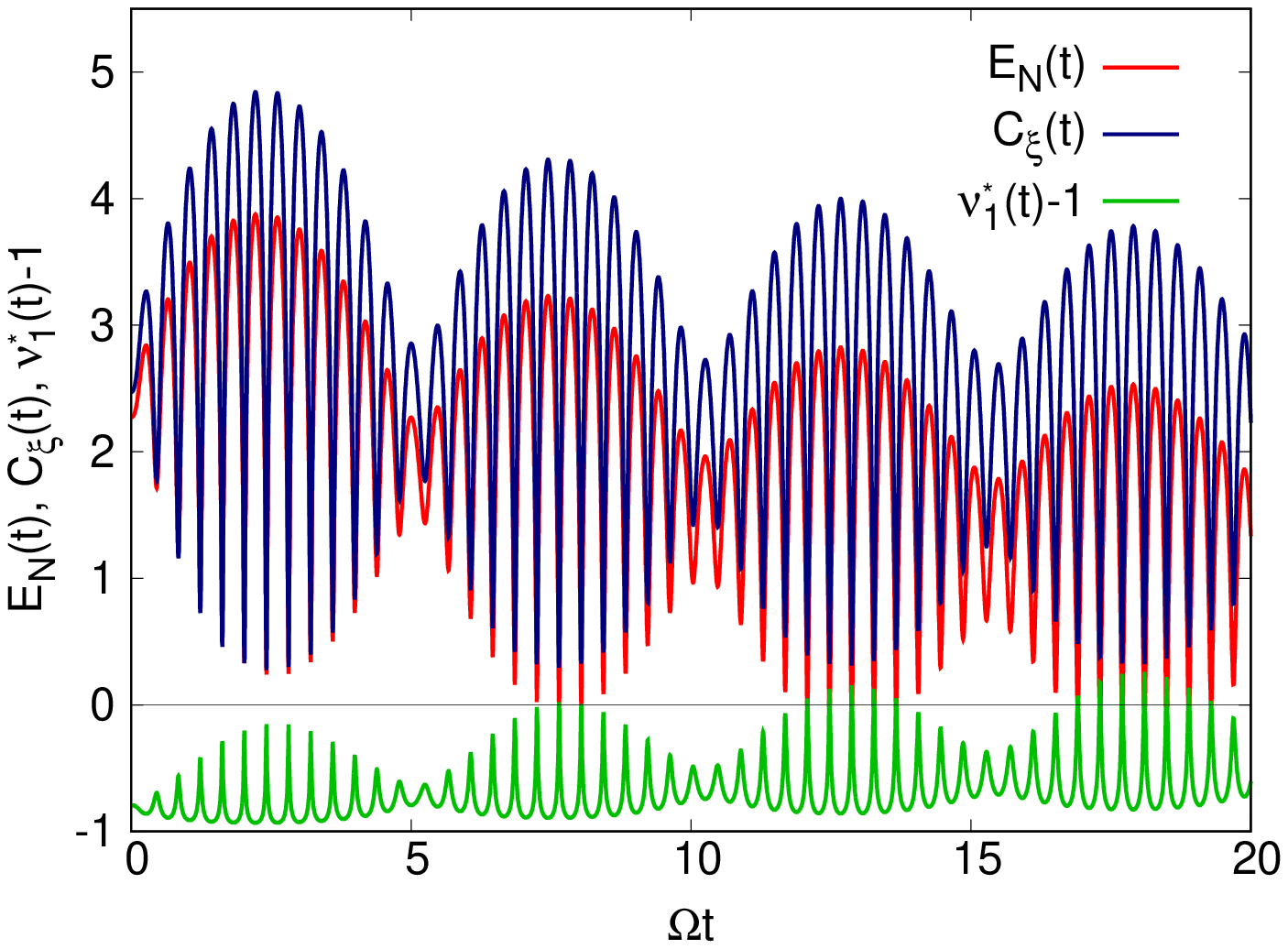} 
\end{center}
\caption{ 
Short time behavior of logarithmic negativity $E_N$,  quantum mutual information $C_\xi$, and the smaller symplectic eigenvalue  $v_1^{_*}$ after partial transpose minus one as a function of time ($p=1.2$, $r_s=1$, $k_B T/(\hbar\Omega)=1$). The coupling constant is $\kappa=5 m \Omega^2$ on the top, and  $\kappa=15 m \Omega^2$ on the bottom sub figures. The color scheme is the same as the concepts in Figure (\ref{fig:entangl_kappa_0}). For better comparison, we draw the zero level with a black line.  All quantities plotted here are taken from the analytical solution of the Markovian HPZ master equation (see Eq.~\eqref{eq:solution}).}
\label{fig:entangl_short}
\end{figure}

\begin{figure}
\begin{center}
\includegraphics[height=0.75\linewidth, angle=270]{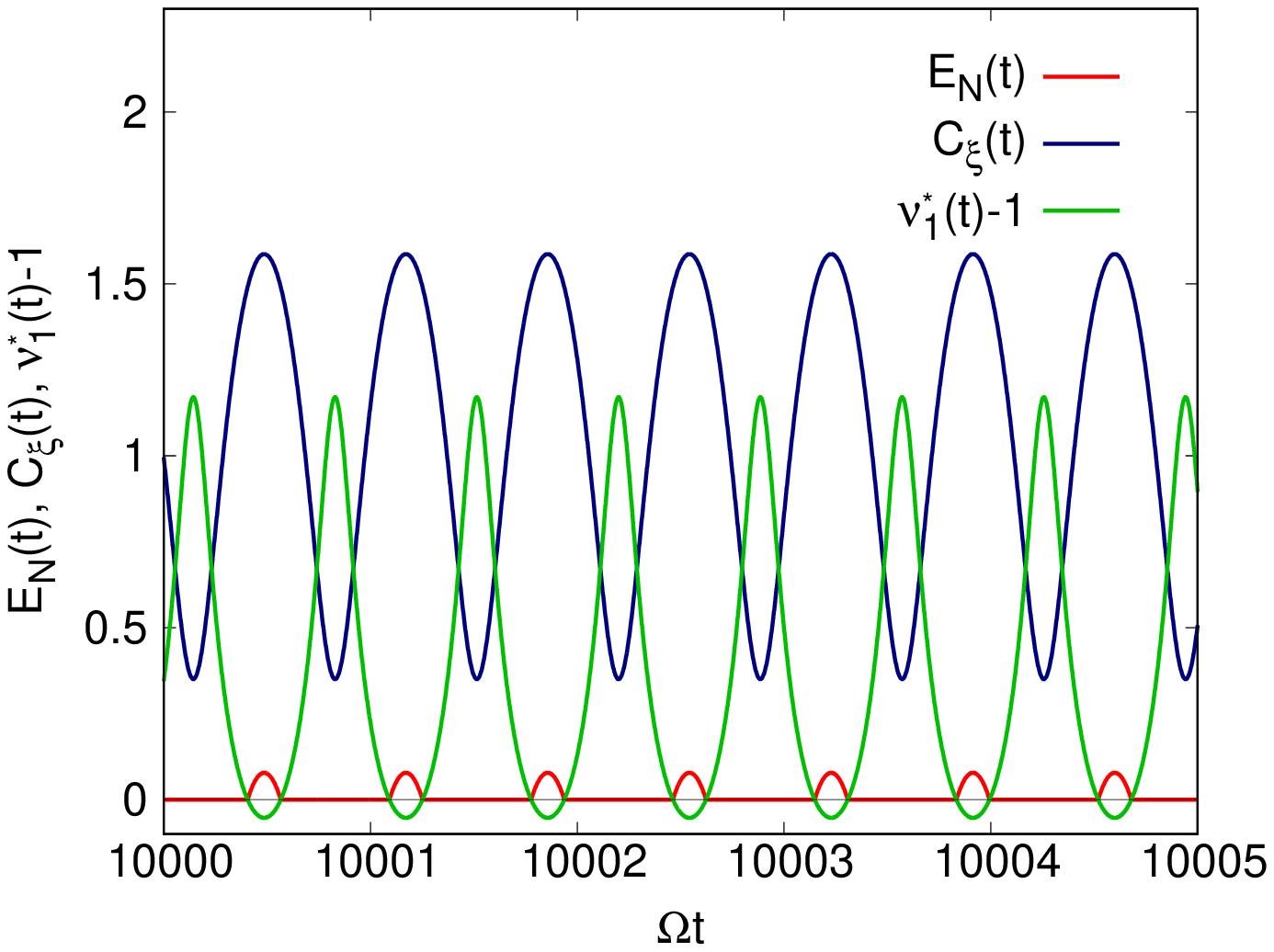} \qquad \includegraphics[height=0.75\linewidth, angle=270]{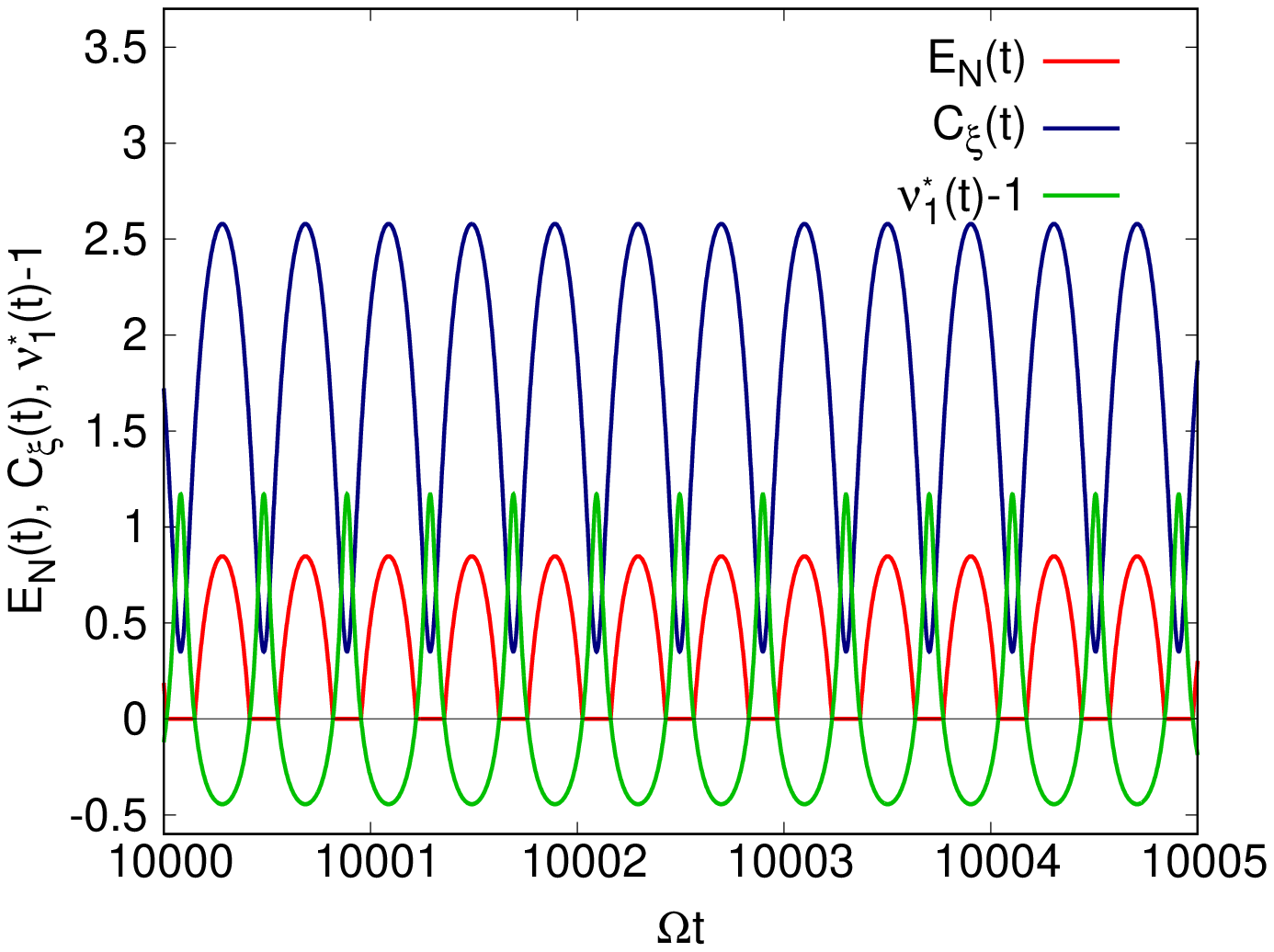}
\end{center}
\caption{ 
Asymptotic behavior of logarithmic negativity  $E_N$,  quantum mutual information $C_\xi$, and the smaller symplectic eigenvalue  $v_1^{_*}$ after partial transpose minus one as a function of time ($p=1.2$, $r_s=1$, $k_B T/(\hbar\Omega)=1$). The coupling constant is $\kappa=5 m \Omega^2$ on the top, and  $\kappa=15 m \Omega^2$ on the right bottom figures. The color scheme is the same as the concepts in Figure (\ref{fig:entangl_kappa_0}). For better comparison, we draw the zero level with a black line.  All quantities plotted here are taken from the analytical solution of the Markovian HPZ master equation (see Eq.~\eqref{eq:solution}).}
\label{fig:entangl_long}
\end{figure}

The two oscillator can be more entangled if the coupling constant $\kappa$ is increased. Such situations for short times are depicted on Fig.~\ref{fig:entangl_short} and in the asymptotic regime in Fig.~\ref{fig:entangl_long} for two values of $\kappa$. These figures show already such a situation when $t_s$ is already infinity: the systems are periodically entangled and disentangled even in the asymptotic regime. The trend is clear: by increasing the direct coupling $\kappa$ the oscillators are more entangled, spending relatively more times in a period belonging to entangled states in the asymptotic regime. The durations of one cycle in Figs~\ref{fig:entangl_long}. are $T_p=\pi/\omega_d$, which can be understood by the following arguments: the asymptotic characteristic function \eqref{eq:asymptotic_in_cm} has only the relevant frequency $\omega_d$, this will also be true for the Gaussian asymptotic covariance matrix, and the quantities $C_\xi(t),E_N(t),\nu_1^*(t)-1$ will oscillate with frequency $2\omega_d$. 

Within one cycle, the relative duration of the entangled or disentangled states can be changed by varying $\kappa$, but also with the squeezing parameter $r_s$ (not shown). It is counterintuitive, but true: at $r_s=0$ the initial covariance matrix describes a separable state in CMR coordinates. Transformation of the covariance matrix \eqref{eq:transformation_of_momenta} is, however, a  symplectic transformation, which does not keep the symplectic spectra after partial transpose according to results \cite{deGosson2021,Serafini}. Thus, even at $r_s=0$ (other parameters are the same as on Fig.~\ref{fig:entangl_long}) the system entangles and disentangles in one cycle.

\begin{figure}
\begin{center}
\includegraphics[height=0.52\linewidth, angle=0]{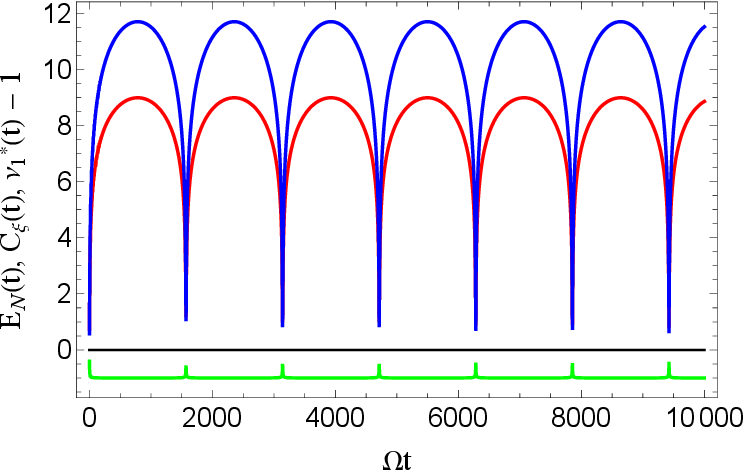} \qquad \includegraphics[height=0.52\linewidth, angle=0]{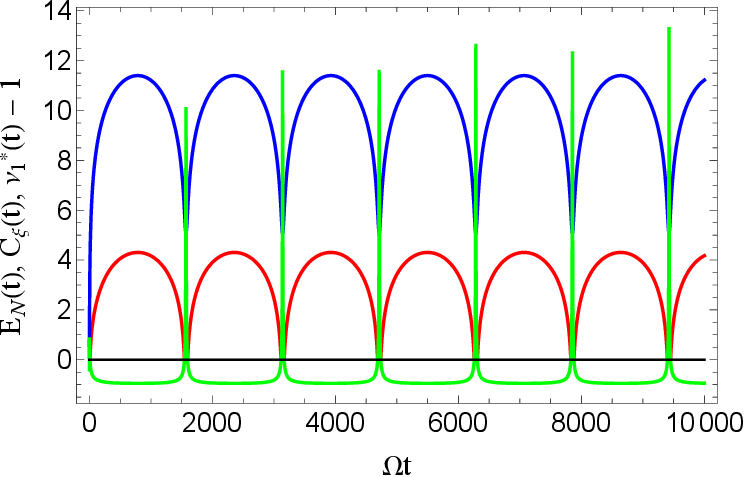}
\end{center}
\caption{ 
Time dependent behavior of logarithmic negativity $E_n$, quantum mutual information $C_\xi$, and the smaller symplectic eigenvalue  $v_1^{_*}$ after partial transpose minus one as a function of time ($p=1.2$, $r_s=0$, $\kappa=-0.249999 m \Omega^2 \rightarrow \omega_d/\Omega=0.002$). The dimensionless temperature  $k_B T/(\hbar\Omega)=0.0001$  on the  upper, and  $k_B T/(\hbar\Omega)=200 $ on the  lower sub figures. The color scheme is the same as the concepts in Figure (\ref{fig:entangl_kappa_0}). For better comparison we draw the zero level with a black line.  All quantities plotted here are taken from the analytical solution of the Markovian HPZ master equation (see Eq.~\eqref{eq:solution}).}
\label{fig:entangl_long_neg}
\end{figure}

The negative coupling range is also an interesting one:
one can keep the physical system consisting of the two oscillators in an entangled state for arbitrarily long time in a wide temperature range, see  Fig.~\ref{fig:entangl_long_neg}, if $\kappa$ is close to $\kappa_\text{crit}=-m\Omega^2/4$. This also means that if $\kappa \to \kappa_\text{crit}$, then $\omega_d \rightarrow 0$ and $T_p \rightarrow \infty$, i.e., the motion  in the relative degree of freedom slows down. At very small temperatures, the system cannot be disentangled (see the upper panel of Fig.~\ref{fig:entangl_long_neg}) and even at very high temperature it can remain in an almost completely entangled state (see the lower panel of the same figure).

Finally, we note that Ref.~\cite{japcsi} (Corollaries 2.26 and 2.27, which extend the Werner-Wolf Theorem \cite{WW} to polynomial Gaussian operators) shows that in a $1+1$-dimensional Gaussian system, if the Gaussian characteristic function represents an entangled density operator, then multiplying it by any polynomial still yields an entangled state.

\section{Summary and conclusions} \label{sec:V}

In this work, we have investigated the Markovian time evolution of a system comprising two harmonic oscillators that are harmonically coupled both to each other and to a common thermal reservoir, modeled as a collection of initially independent harmonic oscillators. We have exploited the fact that this composite system can be effectively reduced to a model. This consists of a single harmonic oscillator; the CM system which is coupled to the thermal bath, and a decoupled free harmonic oscillator: the system of the relative coordinate and momentum \cite{Chou}. We have presented the exact solution of the Markovian Hu-Paz-Zhang (HPZ) master equation, which describes the time evolution of the density operator of the two harmonic oscillators. In addition, we have analyzed the stability conditions associated with the solution. Furthermore, by adopting a Lorentz-Drude type Ohmic spectral density, we have derived the explicit temperature-dependent expressions for the Markovian coefficients of the master equation, which is one of the central outcomes of our study. 

We have analyzed the positivity violations of the master equation and identified nonphysical behavior. In this particular analysis, we have restricted our attention to Gaussian states. The Markovian HPZ master equation emerges as the asymptotic limit of the non-Markovian formulation at long times. Residual nonphysical effects arise because not all density operators are valid as initial states in the Markovian regime. Initially, the non-Markovian HPZ equation accommodates arbitrary density operators, but as the system evolves, it maps them into a restricted subset of physical states. Any attempt to use a density operator outside this subset as an initial condition for the Markovian HPZ equation may lead to nonphysical results. It is worth emphasizing that the Markovian HPZ master equation is derived rigorously from first principles and does not assume any Lindbladian structure. While one possible solution for recovering complete positivity is the inclusion of an additional position diffusion term, the physical origin of this term remains theoretically ambiguous and lacks a universally accepted justification \cite{Dekker, Barnett, Bernad2018}. 

Given the exact time evolution of the density operator under Markovian dynamics, we were able to systematically track several key quantities, including the von Neumann entropy, the  quantum mutual information $C_{\xi}$, the quantity $\nu^*_1-1$, and the logarithmic negativity $E_N$. The latter two are directly associated with the Peres-Horodecki separability criterion. For simplicity, we have employed an  EPR covariance matrix in CMR phase-space coordinates, augmented by a multiplicative prefactor $p$. We have found that setting $p>1$ was essential to avoid immediate violations of the positivity condition, which could otherwise occur at arbitrary temperatures. Since our analysis is restricted to Gaussian states and their Gaussian-preserving dynamics, the Peres-Horodecki criterion is a sufficient and necessary condition for separability. In the considered time evolutions, we have observed a strong correlation between the time dependence of the  quantum mutual information and the logarithmic negativity. Notably, these quantities exhibit similar behavior during intervals where the system remains entangled. This finding is unexpected, given that these two measures capture different physical aspects by definition. 

We have also found that the entanglement properties are strongly temperature-dependent and can be tuned via the coupling constant $\kappa$ and the squeezing parameter $r_s$, both of which may be beneficial for potential applications. Remarkably, even in the absence of direct coupling ($\kappa=0$), the two central oscillators become entangled and subsequently disentangled up to a characteristic time $t_s$. In this case, the bath acts as an effective mediator, maintaining an indirect coupling between the oscillators. We have shown that for times $t>t_s$ the system becomes separable, although its dynamical features continue to exhibit oscillatory behavior. At fixed temperature and for given values of $r_s$ and the multiplicative parameter $p$, increasing the coupling $\kappa$ to positive values results in more rapid oscillations of the asymptotic states and prolongs the duration of entanglement, relative to the periods spent in disentangled states. In the $\kappa>0$ regime, periodic entanglement-disentanglement behavior emerges as a typical asymptotic feature, even at thermal energies on the order of $k_B T \approx \hbar\Omega$. 

A negative coupling in the range $\kappa_\text{crit}<\kappa<0$ at low temperatures makes it possible for the two oscillators to remain entangled for arbitrarily long times. As the temperature increases into the high-temperature regime, i.e., $k_B T>\hbar\Omega_c>\hbar\Omega$, the system remains predominantly entangled over time, with only brief intervals of disentanglement interrupting the dynamics. In all allowed regions of $\kappa$, the asymptotic state exhibits oscillations at the frequency $\omega_d$, with some physical quantities oscillating at integer multiples of $\omega_d$.  These oscillations persist over long timescales. The characteristic function of the asymptotic state in CMR coordinates takes the form of a product: a Gaussian corresponding to the CM degree of freedom, and a limit cycle in the relative coordinates, which depends on the initial state.  

An important aspect of our analysis is that the non-Markovian master equation becomes asymptotically Markovian and, after sufficient time, effectively exhibits Markovian behavior. Within this framework, we have presented a model in which a part of the initially present quantum information can persist despite the presence of external noise. This latter is modeled here with a coupling to a thermal bath. Remarkably, in certain regions of the parameter space, the asymptotic motion in the relative coordinates remains undamped, and the system can retain full entanglement. These findings suggest potential pathways for developing quantum technologies that rely on robust, long-lived entanglement in noisy environments.

\subsection*{Acknowledgments}

We are indebted to Rich\'ard Balka, Andr\'as Bodor,
Andr\'as Frigyik, Tam\'as Kiss, M\'aty\'as Koniorczyk, Nikolett N\'emet, Mikl\'os Pint\'er, Szil\'ard Szalay
and  G\'eza T\'oth for several illuminating conversations and discussions.
G. H. thanks the ”Frontline” Research Excellence Programme of the
NKFIH (Grant no. KKP133827).  A.C., P.A, and J.Z.B. were supported by the Hungarian National Research, Development and Innovation Office within the Quantum Information National Laboratory of Hungary grants no. 2022-2.1.1-NL-2022-00004 and 134437. J.Z.B. acknowledges support from AIDAS-AI, Data Analytics
and Scalable Simulation, which is a Joint Virtual Laboratory gathering the Forschungszentrum Jülich and the French
Alternative Energies and Atomic Energy Commission.

\bibliographystyle{ptephy}
\bibliography{manuscript}

\appendix
\section{Special cases of the Markovian coefficients}
\label{sec:markovian_coeffs}
Temperature independent coefficients $A$ and $B$ have been determined in Ref.\cite{HBCS}
\begin{equation}
A = - 2M\gamma \Omega_c^2\frac{(\Omega_c+z_2+z_3)}{(\Omega_c+z_2)(\Omega_c+z_3)},
\label{eq:A(t)_stac}
\end{equation}
\begin{equation}
B =  \frac{2\gamma \Omega_c^2}{(\Omega_c+z_2)(\Omega_c+z_3)}.
\label{eq:B(t)_stac}
\end{equation}
By using the Vieta's formulae 
 these expressions simplify to \eqref{eq:A_and_B}.

In the weak coupling limit (Eqs. \eqref{eq:a(t)_in_weak_coupling}, \eqref{eq:b(t)_in_weak_coupling}) these coefficients' Markovian limits are:
\begin{eqnarray}
A_w &=& -2 M \gamma \Omega_c^2 \frac{\Omega_c}{\Omega_c^2+\Omega^2}, \label{eq:A_weak} \\
B_w &=&  \frac{2 \gamma\Omega_c^2}{\Omega_c^2+\Omega^2}. \label{eq:B_weak}
\end{eqnarray}

The temperature dependent Markovian coefficients $C$ and $D$ can be calculated along the lines of Ref.\cite{HBCS} at any temperature. Direct calculations lead to

\begin{eqnarray}
C &=&\left({\frac { 2\hbar\gamma \Omega _{c}^{2} \,z_{1}\, \, }{\pi \, \left( 
\Omega _{c}- z_{1}\right)  \left( z_{1}-z_{2} \right)  \left( z
_{1}-z_{3} \right)  }}K(z_1,\nu) + \text{Cycl.} \right) \nonumber\\
&& -\frac{4 \gamma^2 \Omega_c^4k_B T}{z_1(z_1+z_2)(z_1+z_3)(\Omega_c+z_1)(\Omega_c+z_2)(\Omega_c+z_3)} \nonumber \\
&&-\frac{4\hbar\gamma^2 \Omega_c^4\left(z_1^2(z_2+z_3)+\Omega_c(z_1^2+z_2 z_3) \right)}{\pi (z_1^2-z_2^2)(z_1^2-z_3^2)(\Omega_c-z_1)\prod_{i=1}^3(\Omega_c+z_i)}K(z_1,\nu) \nonumber \\
&&+\frac{4\hbar\gamma^2 \Omega_c^4 z_2}{\pi (z_1^2-z_2^2)(z_2-z_3)(\Omega_c^2-z_2^2)(\Omega_c+z_3)}K(z_2,\nu) \nonumber \\
&& +\frac{4\hbar\gamma^2 \Omega_c^4 z_3}{\pi (z_1^2-z_3^2)(z_3-z_2)(\Omega_c^2-z_3^2)(\Omega_c+z_2)}K(z_3,\nu), \label{eqn:C_exact_long}
\end{eqnarray}
and
\begin{eqnarray}
D &=&\left({\frac { 2\hbar\gamma M\Omega _{c}^{2} \,z_{1}^2\, \, }{\pi \, \left( 
\Omega _{c}- z_{1}\right)  \left( z_{1}-z_{2} \right)  \left( z
_{1}-z_{3} \right)  }}K(z_1,\nu) + \text{Cycl.} \right) \nonumber \\
&&+\frac{4 \gamma^2 M\Omega_c^4 k_B T}{(z_1+z_2)(z_1+z_3)(\Omega_c+z_1)(\Omega_c+z_2)(\Omega_c+z_3)} \nonumber \\
&&-\frac{4\hbar\gamma^2 M\Omega_c^4 z_1^2\left( z_1^2+z_2z_3+ \Omega_c(z_2+z_3)\right)}{\pi (z_1^2-z_2^2)(z_1^2-z_3^2)(\Omega_c-z_1)\prod_{i=1}^3(\Omega_c+z_i)}K(z_1,\nu) \nonumber \\
&&+\frac{4\hbar\gamma^2 M\Omega_c^4 z_2^2}{\pi (z_1^2-z_2^2)(z_2-z_3)(\Omega_c^2-z_2^2)(\Omega_c+z_3)}K(z_2,\nu) \nonumber \\
&& +\frac{4\hbar\gamma^2 M\Omega_c^4 z_3^2}{\pi (z_1^2-z_3^2)(z_3-z_2)(\Omega_c^2-z_3^2)(\Omega_c+z_2)}K(z_3,\nu),  \label{eqn:D_exact_long}
\end{eqnarray}
where we have introduced the following notations: $\nu=2 \pi k_B T / \hbar$,  and the function $K(z_i,\nu)$:
\begin{equation}
K(z_i,\nu)=\Psi \left(1-\frac{z_i}{\nu} \right) -\Psi \left( \frac{\Omega_c}{\nu} \right) , \qquad i=1,2,3.
\label{eq:def_of_K(z,nu)}
\end{equation}
Here $\Psi(z)=\Gamma'(z)/\Gamma(z)$ is the digamma function defined in terms of Gamma function $\Gamma(z)$. For the definition of the \text{Cycl.} operator we use Eq.~(54) of Ref.~\cite{HBCS}, i.e.,:
\begin{eqnarray}
f(z_1,z_2,z_3)+ \text{Cycl.} &\equiv& f(z_1,z_2,z_3)+f(z_2,z_3,z_1) \nonumber\\ &&+f(z_3,z_1,z_2). \nonumber
\end{eqnarray}
The first terms in Eqs. (\ref{eqn:C_exact_long}) and  (\ref{eqn:D_exact_long}) in parentheses are from the single integrals in Eqs.~\eqref{eq:def_of_c(t)},\eqref{eq:def_of_d(t)}, others are from triple integrals. Using the Vieta's formulae in (\ref{eqn:C_exact_long}) and (\ref{eqn:D_exact_long}) all those terms greatly simplify to (\ref{eqn:C_exact}) and (\ref{eqn:D_exact}).
The same coefficients in the weak coupling limits \eqref{eq:c(t)_in_weak_coupling}, \eqref{eq:d(t)_in_weak_coupling} reads as
\begin{equation}
C_w=\frac{\hbar \gamma \Omega_c^2}{\pi  \left( 
\Omega^2+\Omega_{c}^2 \right)}\left[ 2\frac{\pi k_{
B} T}{ \hbar\Omega_{c}} + \Psi \left( 1-\frac{i\Omega
\hbar}{2\pi k_{B}T} \right)
+\Psi\left( 1+\frac{i\Omega \hbar}{2\pi k_{B}T}\right)
-2\Psi \left( 1+
\frac{ \hbar\Omega_c}{2\pi k_{B}T} \right) \right],
\label{eqn: C_{2w}}
\end{equation}
and
\begin{equation}\label{eqn: D2w}
D_w=\frac{\hbar M \gamma \Omega_c^2 \Omega \coth{\left( \frac{\hbar \Omega}{2 k_B T}\right)}}{\Omega_c^2+\Omega^2}.
\end{equation}
The first special case of Eqs.\eqref{eqn:C_exact} and \eqref{eqn:D_exact} is when the relations $0 \ll \hbar \Omega \ll \hbar \Omega_c \ll k_B T$  are simultaneously fulfilled. Clearly, this is a classical, high temperature limit. Under these conditions 
\[
K(z_i,\nu) \approx \frac{\nu}{\Omega_c}=2\pi \frac{k_B T}{\hbar \Omega_c},
\]
thus, $C$ and $D$ can be well approximated by
\begin{eqnarray}
C &=& \dfrac{4 \gamma k_B T \Omega_c^2}{\prod_{i=1}^3 (\Omega_c-z_i)}  
\bigg(  1 - \frac{\gamma \Omega_c^2 \big( 2 \Omega_c (\Omega_c-z_1) + (z_1+z_2)(z_1+z_3)    \big)}{z_1(z_1+z_2)^2(z_1+z_3)^2}   \bigg) \nonumber \\
&=& k_B T \frac{z_2 + z_3}{z_1},
\label{eqn: C3}
\end{eqnarray}
and
\begin{eqnarray}
    D &=& \frac{4 \gamma k_B T M \Omega_c^3}{\prod_{i=1}^3 (\Omega_c-z_i)} \bigg(  1 - \frac{\gamma \Omega_c \big(  z_2z_3 - \Omega_c(2\Omega_c-z_1)  \big)}{(z_1+z_2)^2(z_1+z_3)^2}   \bigg)
    \nonumber \\
    &=& - k_B T M (z_2 + z_3).\label{eqn: d3}
\end{eqnarray}
In the classical, but at the same time at the weak coupling limit they are
\begin{equation}\label{eqn: C3w}
C_w = \frac{2 \gamma \Omega_c}{\Omega_c^2 + \Omega^2} k_B T
\end{equation}
and
\begin{equation}\label{eqn: D3w}
 D_w= 2 \gamma M k_B T\frac{\Omega_c^2}{\Omega_c^2 + \Omega^2}.
\end{equation}
For completeness we also enumerate the zero temperature limits of Eqs.
\eqref{eqn:C_exact} and \eqref{eqn:D_exact}, when the approximation
\[
K(z_i,\nu) \simeq\ln\left(-z_i \right) -\ln \left( \Omega_c \right)=\frac{1}{2} \ln\left(z^2_i/\Omega^2_c\right), \qquad T\approx 0, \] 
can be made, provided $\text{Re}\, z_i<0$:
\begin{eqnarray}
C&=&\frac{\hbar(z_2+z_3)(z_1^2+z_2 z_3)}{\pi(z_1-z_2)(z_1-z_3)}K(z_1,0) 
+\frac{\hbar(z_1+z_3)(z_2+z_3)z_2}{\pi (z_1-z_2)(z_3-z_2)}K(z_2,0)
\nonumber \\&&
+\frac{\hbar(z_1+z_2)(z_2+z_3)z_3}{\pi (z_1-z_3)(z_2-z_3)}K(z_3,0),
\label{eqn:C_exact_zero_temp}
\end{eqnarray}
and
\begin{eqnarray}
D&=& \frac{\hbar M z_1^2(z_2+z_3)^2}{\pi(z_1-z_2)(z_1-z_3)}K(z_1,0) 
+\frac{\hbar M(z_1+z_3)(z_2+z_3)z_2^2}{\pi (z_1-z_2)(z_3-z_2)}K(z_2,0) \nonumber \\ 
&&+\frac{\hbar M(z_1+z_2)(z_2+z_3)z_3^2}{\pi (z_1-z_3)(z_2-z_3)}K(z_3,0).
\label{eqn:D_exact_zero_temp}
\end{eqnarray}
These expressions are the same as Eqs. (102) and (103) in Ref.\cite{HBCS}, if one uses Vieta's formulae there.  Finally, the weak coupling limit of these coefficients (see  Eqs. (106) and (107) in Ref.\cite{HBCS}) are 
\begin{equation}
C_w = -\frac{2\hbar\gamma\Omega_c^2}{\pi}\frac{\ln\left(\Omega_c/\Omega\right)}{\Omega_c^2+\Omega^2}, \qquad T=0,
\label{eq:C_w_asymptotic} 
\end{equation}
and
\begin{equation}
D_w = \frac{\hbar M\gamma\Omega_c^2 \Omega}{\Omega_c^2+\Omega^2}, \qquad T=0. 
\label{eq:D_w_asymptotic}
\end{equation}
In  Table \ref{tab: Table 1}., we summarize the different limiting forms of the master equation's Markovian coefficients.

\end{document}